%% file: manuscript.tex
\documentclass[journal]{IEEEtran}
\IEEEoverridecommandlockouts
\usepackage{cite}
\usepackage{tikz}

\input{packages}

\input{newcommands}

\begin{document}

\title{
  \input{title}
}

\author{
\IEEEauthorblockN{Claudio Cicconetti}
\IEEEauthorblockA{\textit{IIT, CNR} --
Pisa, Italy \\
c.cicconetti@iit.cnr.it}
\and
\IEEEauthorblockN{Marco Conti}
\IEEEauthorblockA{\textit{IIT, CNR} --
Pisa, Italy \\
m.conti@iit.cnr.it}
\and
\IEEEauthorblockN{Andrea Passarella}
\IEEEauthorblockA{\textit{IIT, CNR} --
Pisa, Italy \\
a.passarella@iit.cnr.it}
}

\author{Claudio~Cicconetti,
        Marco~Conti,
        and Andrea~Passarella%
\IEEEcompsocitemizethanks{\IEEEcompsocthanksitem All the authors are with the Institute of Informatics and Telematics (IIT) of the National Research Council (CNR), Pisa, Italy.}%
}

\IEEEtitleabstractindextext{%
\begin{abstract}
  \input{abstract}
\end{abstract}

}

\maketitle

\begin{tikzpicture}[remember picture,overlay]
\node[anchor=south,yshift=10pt] at (current page.south) {\fbox{\parbox{\dimexpr\textwidth-\fboxsep-\fboxrule\relax}{
  \footnotesize{
     \copyright 2021 IEEE.  Personal use of this material is permitted.  Permission from IEEE must be obtained for all other uses, in any current or future media, including reprinting/republishing this material for advertising or promotional purposes, creating new collective works, for resale or redistribution to servers or lists, or reuse of any copyrighted component of this work in other works.
  }
}}};
\end{tikzpicture}%

\IEEEdisplaynontitleabstractindextext

\IEEEpeerreviewmaketitle

\bibliographystyle{plain}


\input{main}

\end{document}

%% file: packages.tex
\usepackage{amsmath,amssymb,amsfonts}
\usepackage{algorithmic}
\usepackage{graphicx}
\usepackage{textcomp}
\usepackage{xcolor}
\usepackage[ruled]{algorithm2e} 
\usepackage[nolist]{acronym}
\usepackage{soul}
\usepackage{marginnote}
\usepackage{url}
\usepackage[inline]{enumitem}
\usepackage{tabularx}
\usepackage{amsthm}

%% file: newcommands.tex
\newcommand{\myssec}[2]{%
  \subsection{#1}%
  \label{sec:#2}%
}
\newcommand{\rsec}[1]{%
  Sec.~\ref{sec:#1}%
}

\newcommand{\added}[1]{%
  {\color{blue}#1%
  }%
}
\renewcommand{\added}[1]{#1}
\newcommand{\removed}[1]{%
  {\color{red}#1%
  }%
}
\renewcommand{\removed}[1]{}

\newcommand{\myfigeps}[3][width=3.1in]{%
\begin{figure}[t]%
\centering%
\includegraphics[#1]{figures/#2}%
\vspace{-1em}%
\caption{#3}%
\label{fig:#2}%
\vspace{-1.5em}%
\end{figure}%
}

\newcommand{\rfig}[1]{Fig.~\ref{fig:#1}}

\newcommand{\rtab}[1]{Table~\ref{tab:#1}}
\newcommand{\req}[1]{Eq.~(\ref{eq:#1})}
\newtheorem{lemma}{Lemma}

\newenvironment{myinlinelist}%
{%
\begin{enumerate*}[label=(\roman*)]%
}%
{%
\end{enumerate*}%
}

{%
\begin{itemize}[parsep=0em,leftmargin=*,label={--}]%
}%
{%
\end{itemize}%
}

{%
\begin{enumerate}[parsep=0em,leftmargin=*,label=\arabic*.]%
}%
{%
\end{enumerate}%
}

%% file: title.tex
A Decentralized Framework for Serverless Edge Computing in the Internet of Things

%% file: abstract.tex
Serverless computing is becoming widely adopted among cloud providers,
thus making increasingly popular the Function-as-a-Service (FaaS)
programming model, where the developers realize services by packaging
sequences of stateless function calls.
The current technologies are very well suited to data centers, but
cannot provide equally good performance in decentralized environments,
such as edge computing systems, which are expected to be typical
for Internet of Things (IoT) applications.
In this paper, we fill this gap by proposing a framework for efficient
dispatching of stateless tasks to in-network executors so as to
minimize the response times while exhibiting short- and long-term
fairness, also leveraging information from a virtualized network
infrastructure when available.
Our solution is shown to be simple enough to be installed on
devices with limited computational capabilities, such as IoT gateways,
especially when using a hierarchical forwarding extension.
We evaluate the proposed platform by means of extensive emulation
experiments with a prototype implementation in realistic conditions.
The results show that it is able to smoothly adapt to the mobility of
clients and to the variations of their service request patterns,
while coping promptly with network congestion.

%% file: keywords.tex
Internet of Things services, Software-defined networking, Overlay networks, Computer simulation experiments

%% file: main.tex
%
  \section{Introduction}%
  \label{sec:introduction}%
  \input{introduction}
  \section{State of the art}%
  \label{sec:soa}%
  \input{soa}  \section{Architecture}%
  \label{sec:contribution}%
  \input{contribution}
  \section{E-router algorithms}%
  \label{sec:algorithms}%
  \input{algorithms}%

%
  \section{Scalability extension}%
  \label{sec:scalability}%
  \input{scalability}%


%
  \section{Performance evaluation}%
  \label{sec:eval}%
  \input{eval}%

%
  \section{Conclusions}%
  \label{sec:conclusions}%
  \input{conclusions}
\bibliography{%
bib/edgecomputing,%
bib/5g,%
bib/common,%
bib/claudio%
}


\input{biographies}


\input{acronyms}

%% file: introduction.tex
\added{%
\IEEEPARstart{E}{dge} computing is a consolidated type of network
deployment where the execution of services is pushed as close as
possible to the user devices.
This allows to achieve faster response times, relieve user devices
from computationally-intensive tasks, reduce energy consumption,
and lower traffic requirements, compared to current solutions where
applications run fully on user devices or in a remote data
center~\cite{Campbell2019}.
The \ac{IoT} and mobile network domains are expected to gain maximum
benefit from edge computing~\cite{Pan2017,Premsankar2018}, for which
tangible commitments from industry are the reference architecture
published by the OpenFog
Consortium~\cite{OpenFogConsortiumArchitectureWorkingGroup2017} and
the interfaces and protocols for an inter-operable \ac{MEC}
standardized within \ac{ETSI}~\cite{Taleb2017a}.

Unrelated from edge computing, \textit{serverless} is emerging among
the cloud technologies to hide completely to the developer the
notion of an underlying server (hence its name) and provide a true
pay-per-use model with fine
granularity~\cite{Castro:2019:RSC:3372896.3368454}.
To deliver this promise, a serverless platform must provide infinite
scalability, achieved through a flexible up-/down-scaling of lean
virtualisation abstractions (typically, containers).
This has led to the widespread adoption of the \ac{FaaS} programming
model, where the application does not run as a whole process, but
rather it is split into short-lived stateless tasks that can be
executed by different workers without a tight
coordination~\cite{Hendrickson2016}.
It is expected that serverless computing will become widespread in
the future~\cite{Khandelwal2020}.
The seamless adoption of a \ac{FaaS} paradigm in an edge system
would allow to re-use applications, on both the provider's and the
user's side, which were designed originally for serverless computing
and are growing significantly in number and importance~\cite{Nastic2017}.
}

\myfigeps[width=3.3in]{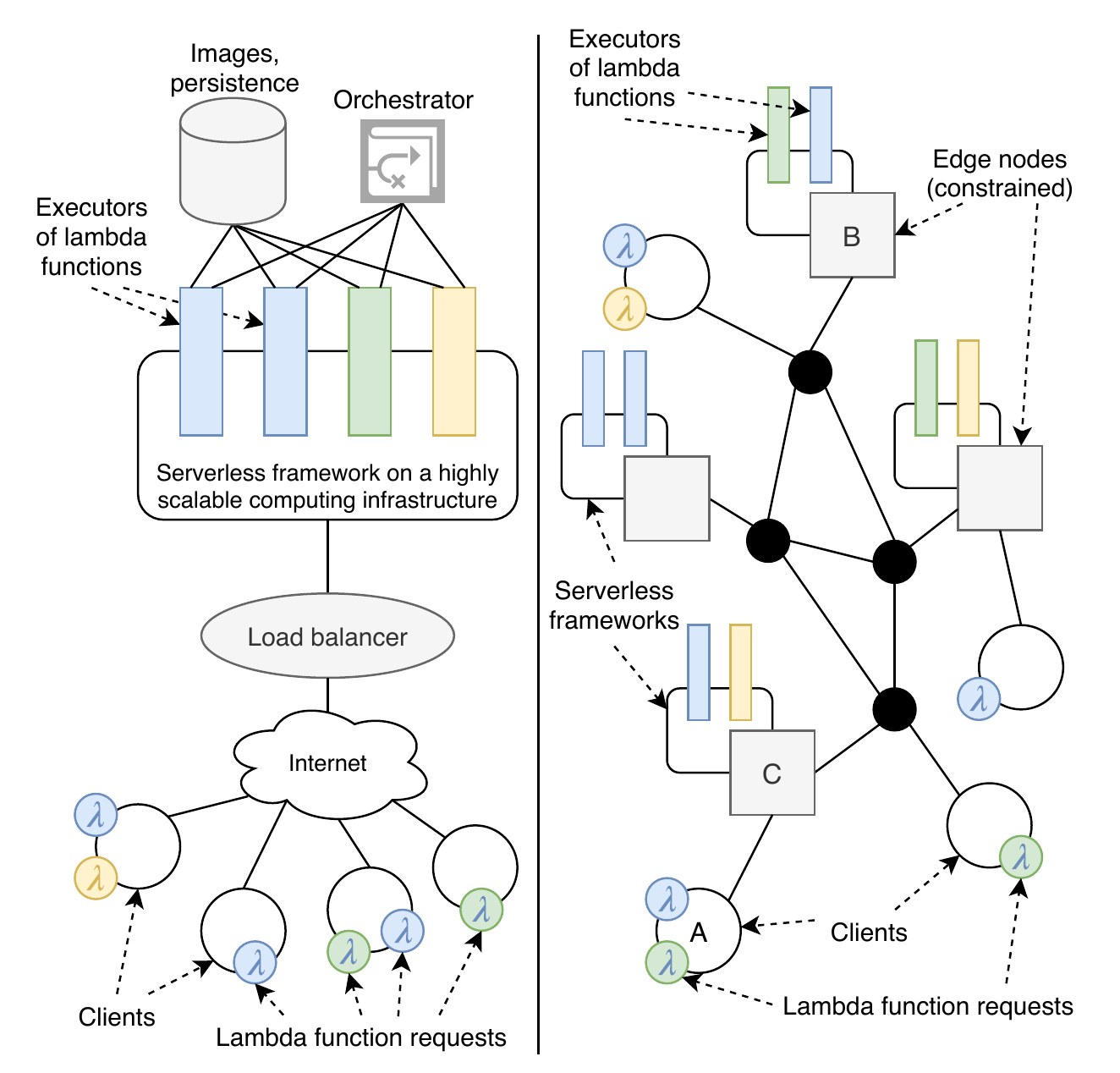}{%
Serverless computing on a distributed (left) vs.\ edge (right) environment.}

However, a straightforward deployment of serverless frameworks in
an edge computing domain may not always be possible because of their
different characteristics~\cite{Glikson2017}.
We elaborate on this point in the following with the help of the
examples in \rfig{distr-vs-edge}.
On the left part, we show a typical (simplified) set up of a
serverless system, inspired from the architecture of Apache
OpenWhisk\footnote{\url{https://openwhisk.apache.org/}}.
Generally, the core of the framework is a flexible container-based
virtualization system hosted in a data center (i.e., operated by a
single entity and running on high-end servers with fast connectivity),
which is controlled by an orchestrator so that the number of
\textit{executors} matches the current demands from clients.
The executors are the applications that respond to the requests of
task executions, also called \textit{lambda functions}.
A load balancer at the ingress of the serverless system dispatches
the incoming lambda functions towards the executors currently
deployed.
A database is represented in the figure to offer a repository of
container images to be dynamically deployed, a persistence service,
user credentials, etc.
Overall, serverless systems can be considered viable implementations
of many \textit{distributed computing} models, such as \ac{BSP} as
elaborated in~\cite{Jonas2017}.


In this work we focus instead on a different scenario where:
\begin{myinlinelist}
  \item the tasks are executed by devices (\textit{edge nodes})
  with limited computation capabilities, such as WiFi access points
  or IoT gateways;

  \item these devices have limited connectivity towards a remote
  data center, which makes it impractical to realize \textit{vertical
  offloading}, i.e., delegation to the cloud of the execution of
  tasks currently exceeding the capacity of an individual edge node;

  \item the clients are interspersed with edge nodes, i.e., there is
  no single point of access for all the users in the system.
\end{myinlinelist}
An example of our target system is illustrated in the right part
of \rfig{distr-vs-edge} and, in its essence, is also the subject
of investigation in~\cite{8662815}, where the authors envision that
every edge node (or small cluster of edge nodes) hosts a serverless
platform with state-of-the-art technology (e.g., every cluster may
run an OpenWhisk instance).
This allows to realize \textit{horizontal offloading}, where
delegation of task execution happens from one serverless platform
to another in the edge network, thus achieving resource pooling at
the edge, without resorting to the cloud.
In this work we embrace this vision, and focus on the specific issue
of distributing lambda function requests from clients to executors
on the various serverless platforms available, which can be also
heterogeneous in hardware/software and owned by different entities.
%
%
%
%
%

Note that this operation is different from the orchestration of
multiple executors in a flexible computation infrastructure, as in
a traditional serverless platform on cloud resources.
%
%
First, despite some initial enthusiastic experiments with container
clusters on constrained devices, such as K8s on Raspberry
Pi's~\cite{10.1007/978-981-13-1056-0_66}, the high overhead and the
orchestration system being designed for a very different target
make such deployments unsuitable for a production system, as
demonstrated in~\cite{8903766} for real-time 5G applications and
in~\cite{234779} for edge AI.
Second, the lack of a natural point for the installation of a
system-wide collector makes it inefficient to use a standard load
balancer, such as NGINX\footnote{\url{https://www.nginx.com/}},
unless the edge operator is willing to pay the price of traversing
multiple times the same link: e.g., with reference to \rfig{distr-vs-edge},
if the load balancer is installed in B, then lambda requests from
A executed on C would have to unnecessarily traverse twice all the
links between B and C, which could be unacceptable in many cases.

To overcome these limitations, \added{in \cite{Cicconetti2018} we
have first proposed a \textit{decentralized} architecture, as
opposed to the \textit{distributed} approach of traditional serverless
frameworks, which is illustrated for completeness in \rsec{contribution}.}
%
Briefly, the edge nodes providing user devices with network access
(which we call edge routers --- \textit{e-routers}, defined in
\rsec{contribution}) take autonomous decisions on where to forward
the incoming lambda request, among the possible executors of matching
type (which we call edge computers --- \textit{e-computers}, defined
in \rsec{contribution}).
To achieve fast processing rate on e-routers with limited capabilities
we split this decision into two sub-functions, called weight updating
and destination selection, by analogy with routing in \ac{IP}
networks, which happens autonomously at each router (fast decision,
called forwarding), but exploits cost information collected from
routing protocols at much longer time scales.

\added{In the same work~\cite{Cicconetti2018}, we have also outlined
algorithms for the realization of these sub-functions}.
\textit{Weight updating} is done by each e-router by measuring a
smoothed average of every lambda function response time towards any
e-computer, and avoiding paths indicated as congested by a \ac{SDN}
controller supervising the network layer \added{(see \rsec{algorithms:weight})}.
%
%
For \textit{destination selection}, we consider three policies
(well-known in the literature) to show that our framework is flexible
enough to incorporate algorithms with different objective functions
to fit specific use cases\added{, which are evaluated according tot their
short- and long-term fairness in \rsec{algorithms:destination}}.
%
We assess the scalability of our solution in~\rsec{scalability}.
Specifically, by means of testbed experiments, we show that performance
may be impaired as the number of services in the edge computing
domain grows, but overcome this issue by enabling a hierarchical
structure: an e-router may appear as an e-computer to other
e-routers so as to receive lambda function requests, which it dispatches
to nearby e-computers that were not addressable directly by other e-routers.
%

Finally, \added{a comprehensive evaluation is carried out in an
emulated environment}, also integrating OpenWhisk platforms.
The results, reported in~\rsec{eval}, show that:
\begin{enumerate*}[label=(\roman*)]
  \item the interaction with the \ac{SDN} controller is beneficial
  to relieve user applications from network congestion;

  \item our decentralized architecture
  is effective in following load variations in several conditions,
  and can achieve almost same performance as a \added{distributed
  system} with equivalent aggregate capacity;

  \item the system can achieve scalability, even with low-end edge
  nodes, by means of hierarchical forwarding.
\end{enumerate*}
Our framework is available as open
source\footnote{\url{https://github.com/ccicconetti/serverlessonedge}}.

%% file: soa.tex
The adoption of the serverless and \ac{FaaS} paradigms in an
edge environment has been first proposed by Baresi
\textit{et.~al}, who have analyzed the issues and proposed
high-level directions in their works.
Among them, \cite{Baresi2019} is particularly relevant.
In addition to providing a convincing motivation, which is fully
aligned with our view expressed in \rsec{introduction}, they envision
different types of deployment, all with multiple serverless platforms
scattered in the network, depending on specific \ac{IoT} use cases.
Furthermore, they provide a detailed architecture of such platforms,
supported by a prototype realized with OpenWhisk and other open
source products.
In this work we complement this result by defining a framework for
the distribution of stateless tasks among the platforms, which
has remained so far unspecified.

In a more general sense, the topic of edge computing has been already
investigated broadly in the scientific literature, see for instance
the survey~\cite{Mach2017}.
In the following we report the state-of-the-art contributions that are
most relevant to the topic addressed in this work.

On the one hand, solutions to the so-called \textbf{edge-cloud
problem} have been proposed: how to dynamically dispatch jobs to a
network of edge nodes based on the application requirements and the
current load on the servers doing the computations~\cite{Tan2017},
\cite{Filip2018}, \added{\cite{Aral2019}, \cite{Chen2018d}}, thus
realizing vertical offloading (as defined in \rsec{introduction}).
%
%
\added{%
Different from our work, these architectural approaches assume that
(i) either the edge nodes are a geographic extension of the data
centers, hence they are as much stable and dependable, and deployed
in limited numbers, or (ii) the offloaded tasks are long-lived,
thus it is suitable to assign tasks to edge nodes by solving a
system-wide optimisation problem.
}
Furthermore, they do not exploit integration with underlying \ac{SDN}
facilities, which is instead generally considered beneficial and
it has been investigated in several research works, see for instance
the survey in~\cite{Baktir2017}.
%
%
%
On the other hand, several works have proposed a \textbf{lightweight
orchestration}, by scaling down cloud-oriented paradigms to less
powerful servers and faster dynamics.
Examples include Picasso~\cite{Lertsinsrubtavee2017}, from which
we reuse the concept of providing the applications with an \ac{API}
whose routines are executed by the network in a transparent manner
to clients, and \textit{foglets}~\cite{Saurez2016}, based on
containers for an easier and faster migration of functions based
on situation-awareness schemes.
Both studies put forward efficient ways to periodically tune the
deployment of \acp{VM}/containers in edge computers, which are
out of the scope of the current work.
On the other hand, some recent works have recently revamped the
topic of distributed scheduling of ``tasklets'' ($\simeq$~\textit{lambdas})
to realize complex applications with shared computational resources.
In~\cite{Edinger2017} the authors propose a way to reduce the number
of tasklet execution failures, which stems from the high unreliability
of the users offering computational resources.
This is not applicable to the case under study where the edge nodes
can safely be assumed to disconnect or fail very sporadically.
Instead, a middleware is put forward in~\cite{7568580} to orchestrate
tasklets through brokers, thus building a hybrid peer-to-peer
network; however, the architecture is completely unaware of the
underlying network and, in particular, does not collaborate with
\ac{SDN} functions as we propose.

With specific reference to \textbf{edge computing} environments,
in~\cite{Nastic2018} the authors propose an abstraction called
\textit{Software-Defined Gateway}, which hides the complexity of
local \ac{IoT} devices and offers a clean \ac{API} for the integration
with widely adopted serverless platforms; they focus on the DevOps
aspects, and mention the issue of ``scheduling [the] functions
execution on loosely coupled and scarce edge resources'' as an open
research issue, which we address in this work.
In~\cite{Destounis2016} the authors take a more radical approach
and assume that every edge node may decide at the same time where to
route every incoming task (at a network-level) and which node has
to execute the remaining processing, where applications are assumed
to be modeled as directed computation graphs.
In this setting they find theoretical results on the system stability,
provided that the edge nodes employ a form of back-pressure, i.e.,
basically they take distributed decisions based on the length of
their backlog queues.
The approach is interesting and, even though we could not reuse the
same exact methodology because of the different assumptions about
the target system, we have inspired from the results
in~\cite{Destounis2016} to define the \textit{Round-robin} algorithm,
defined and analyzed in \rsec{algorithms:destination}.
Furthermore, in~\cite{Cicconetti2019} we have proposed a solution to
estimate the lambda execution time using real-time measurements
from the executors, but this technique cannot be employed when
e-routers are hosted on low-power \ac{IoT} gateways due to the
relatively high memory and computation requirements.
\added{Finally, in~\cite{Aral2018} the authors investigate the issue of
decentralised replica placement at the edge, which is relevant
to the activation phase of executors on edge nodes, not studied in
this work but currently under investigation.}

%
%
%
%
%

This paper is an extended version of~\cite{Cicconetti2018}.
New material includes:
the analysis of the destination selection algorithms in~\rsec{algorithms},
the scalability extension in~\rsec{scalability}, the entire performance
evaluation in \rsec{eval}, also integrating OpenWhisk systems.

%% file: contribution.tex
In this section we describe our proposed architecture for \ac{FaaS}
in an edge system.
%
%
%
As introduced in~\rsec{introduction}, we assume that \ac{IoT} devices
consume serverless services from edge nodes in an \textit{edge
computing domain} consisting of:
\begin{enumerate*}[label=(\roman*)]
  \item edge computers (\textit{e-computers}), i.e., networking
  devices offering their computational capabilities for the execution
  of lambda functions requested by clients;

  \item edge routers (\textit{e-routers}), i.e., networking devices
  that handle lambda execution requests from clients by injecting
  them into the edge computing domain and forwarding the
  respective responses from the e-computers to the clients.
\end{enumerate*}
Any edge node may play both roles.
Moreover, there may be plain networking devices enabling communication
between edge nodes.

\myfigeps{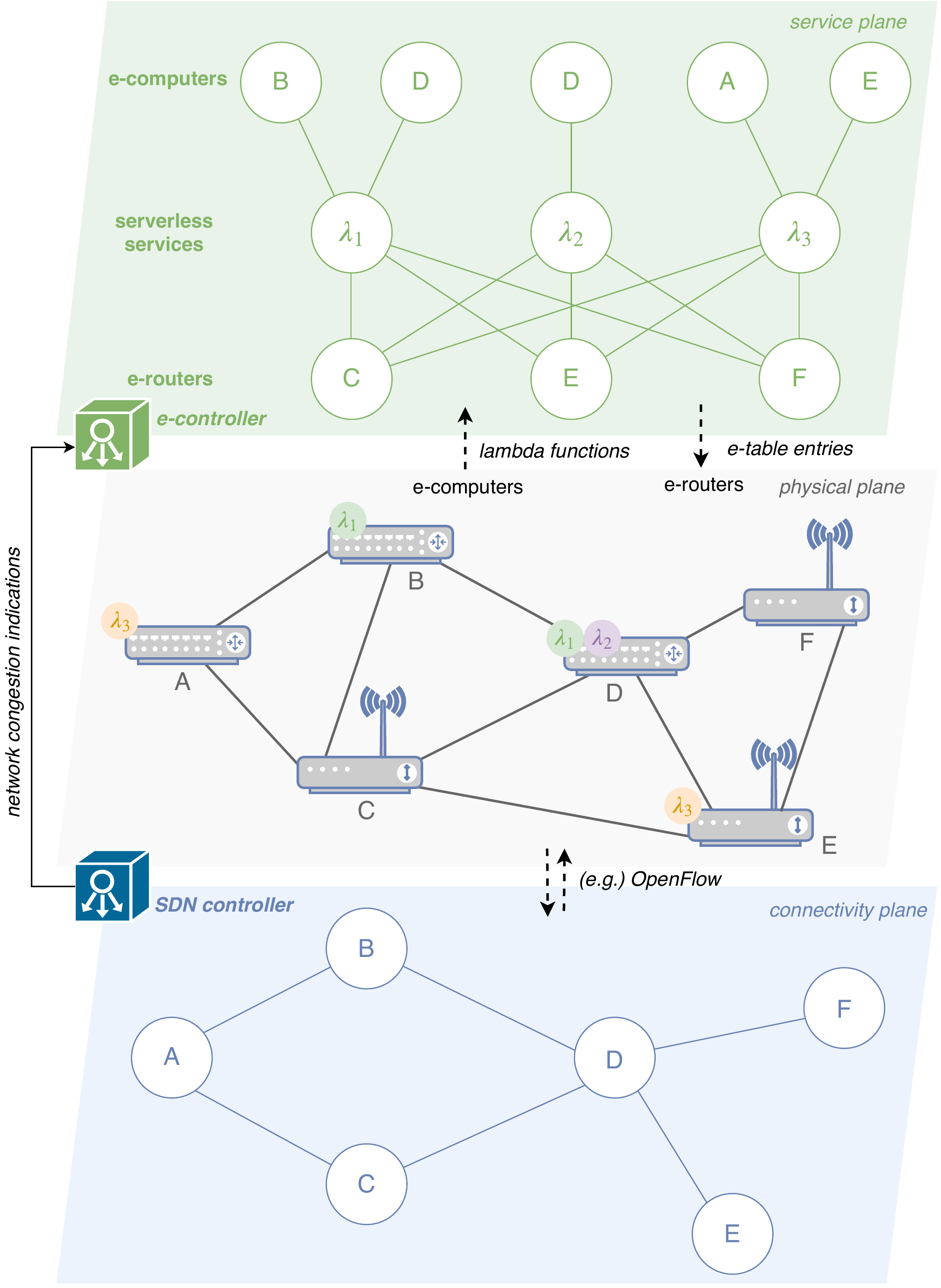}{%
System model example, showing the service plane (as perceived by
the e-controller) and the connectivity plane (as perceived by the
\ac{SDN} controller).}

The complete system model is illustrated by means of the example
in~\rfig{arch-layered}, also showing:
\begin{enumerate*}[label=(\roman*)]
  \setcounter{enumi}{2}

  \item the \textit{\ac{SDN} controller}, which has responsibility
  for maintaining connectivity in the network and provides
  up-to-date topology and network congestion information;

  \item the edge controller (\textit{e-controller}), which discovers
  the capabilities of the e-computers in the domain and configures
  accordingly the e-routers so that they can set up and maintain
  their e-router forwarding tables (\textit{e-tables}).
\end{enumerate*}

Therefore, to make the forwarding process of e-routers reactive to
the fast changing conditions in an edge system
we introduce the concept of \textbf{weight}, which is assigned by
e-router $s$ to every destination e-computer $d$ and lambda function
$i$, and denoted as $w^s_{i,d}$ in the following.
%
%
Such weight is a measure of the \textit{cost} to execute the
associated lambda on the target e-computer from the point of view
of a given e-router: the lower the cost, the more desirable the
destination.
Weights are adapted over time by every e-router independently, so
that this process \added{is lightweight} and may follow closely the
load/application variations on a time scale that is intermediate
between the add/removal of entries in the e-tables and the per-lambda
destination selection.


Our definition of e-tables in the e-routers basically establishes
an overlay model of an edge domain for serverless computing, as
illustrated in the top part of~\rfig{arch-layered}.
At a steady state, the overlay looks the same to all
e-routers\footnote{This will be revisited in~\rsec{scalability} for
scalability reasons.}, but the weights associated to the same lambda
and destination may be different on each e-router, since they depend
on the connectivity of the e-routers to the respective e-computers
and the past history of executions experienced.
In this work we tackle the following research challenges: how to
keep the weights updated in the e-tables over time and which
destination to select for every incoming lambda request (\rsec{algorithms})
and how to achieve scalability as the size of e-tables increases
(\rsec{scalability}).

Finally, we note that our solution gracefully fits into the \ac{ETSI}
\ac{MEC} reference architecture, as shown in the supplementary
material.

%% file: algorithms.tex
In this section we propose algorithms for the realization of an
efficient execution of serverless functions in the architecture
illustrated in~\rsec{contribution}, separately for the two sub-problems
of weight updating and destination selection described therein.

\myssec{Weight updating}{algorithms:weight}

From~\rsec{contribution}, the weight updating problem at e-router
$s$ for lambda $i$ consists in deciding how to assign $w^s_{i,d}$
so that this value reflects a smaller cost incurred by a client if
the function is executed by computer $d$ than that paid if it was
executed by computer $d'$, where $w^s_{i,d} < w^s_{i,d'}$.
It is clear that, from a design perspective, weight updating depends
on the objective function that we wish to pursue.
In principle, there are several goals that one may pursue when
designing a system for edge computing: energy efficiency, \ac{IP}
traffic balancing, user application latency reduction.
While all objective functions have some relevance in the big picture,
we believe the latter, i.e., minimization of latency, will be one
of the key drivers for edge computing adoption, and hence it is
considered in this work as the chief performance goal to be achieved.
In fact, we aim at setting $w^s_{i,d}$ equal to the expected
latency incurred by e-router $s$ if forwarding lambda function $i$
to the e-computer $d$.
%
%

There are several components that contribute to such latency:
%
%
%
%
%
%
%
\begin{enumerate*}[label=(\roman*)]
  \item \textit{processing delay} at the e-computer, both directly
  due to the execution of $\lambda_i$ itself and indirectly caused
  by other concurrent tasks;

  \item \textit{transfer delay}, including the hop-by-hop transmission
  of data from the client to the e-router to the e-computer and
  back to the client, any transport protocol overhead, and queuing
  at all intermediate routers and switches;

  \item \textit{application queuing delay}, which may happen if the
  e-computer has a pre-allocated number of workers per \ac{VM}/container.
\end{enumerate*}
Furthermore, we expect the components above to be highly variable
over time and difficult to predict in most use cases of practical
interest for \ac{IoT}.
%
%
Therefore, static manual configuration of weights based on off-line
analysis is not a viable option.
Instead, we consider a common smoothed average rule for weight
updating.

Specifically, each e-router $i$ performs an independent update of
the weight $w^i_{j,d}$ associated to lambda $\lambda_j$ directed
to destination $d$:
\begin{equation}
w^i_{j,d}(t^+) = \begin{cases}
\infty & \text{if net cong} \\
\alpha \cdot w^i_{j,d}(t^-) + (1 - \alpha) \cdot \delta_{j,d}(t^+) & \text{otherwise}
\end{cases}
\label{eq:weightupdate}
\end{equation}
where $\delta_{j,d}(t^+)$ is the latency measured at current time
$t^+$ for $\lambda_j$ executed by e-computer $d$, $w^i_{j,d}(t^-)$
is the weight's previous value (if no such value is available then
it is $w^i_{j,d}(t^+) = \delta_{j,d}(t^+)$), $\alpha$ is a smoothing
factor, and the condition ``if net cong'' (= if there is temporary
network congestion along the path from e-router $i$ to e-computer
$d$) is indicated by the e-controller to the e-router $i$ as needed.

Such congestion cases may be detected by the \ac{SDN} controller
using any existing network monitoring mechanism and conveyed to the
e-controller, which will therefore proactively disable the corresponding
entry on the e-routers to avoid undesirable delays and exacerbating
congestion further (first condition in \req{weightupdate}).
As the congestion situation is resolved, either due to decreased
demands or because of a reconfiguration of the network paths, the
weight will be restored to the last value it had before congestion.
On the other hand, under normal network conditions, the second
condition in \req{weightupdate} will continuously keep the weight
aligned with the average overall latency, without distinguishing
on whether it is due to processing or transfer.
This is consistent with the objective of minimizing latency as
a user-centric, rather than network- or operator-centric, metric.

\myssec{Destination selection}{algorithms:destination}

The destination selection problem consists in finding at e-router
$s$ the \textit{best} e-computer $\bar{d}$ where an incoming lambda
$i$ request should be forwarded, among the possible destinations
$d \in \mathcal{D}_i$, each assigned a weight $w^s_{i,d}$.
Since the weight, according to the update procedure
in~\rsec{algorithms:weight}, is a running estimate of the latency,
a trivial solution to the destination selection problem could be
to always select the destination with the smallest weight.
However, the same algorithm is used by all the e-routers in the
edge computing domain and, as the reader may recall
from~\rsec{contribution}, by design the e-routers do not communicate
with one another (either directly or through the e-controller), to
keep the process lightweight and scalable to a high number of
e-routers.
Therefore, we argue that such a simplistic approach could, in
principle, be inefficient because of undesirable \textit{herd
effects}, which are well known in the literature of scheduling in
distributed systems~\cite{Roussopoulos2006}.
Briefly, if all e-routers agree that e-computer $x$ is the best one
to serve a given lambda $i$, then all will send incoming requests
to it, thus overloading it, which eventually will result in performance
degradation, which will make all e-routers move together towards
the former second-best e-computer $y$, which will then become
overloaded, and so on.
To capture the need for a compromise between selfish minimization
of delay and fair utilisation of resources, we redefine the
well-established notion of proportional fairness (e.g.,~\cite{Kelly1998})
as follows: \textit{a destination selection algorithm is fair if
any destination is selected a number of times that is inversely
proportional to its weight.}
In other words, this means that a ``proportional fair'' destination
selection algorithm tends to select (proportionally) more frequently
a destination with a lower weight, i.e., latency estimation.
This notion also embeds an implicit assumption of load balancing
if the dominant effect on latency is the processing time.

In the following we propose three algorithms that exhibit an
incremental amount of proportional fairness, ranging from none to
short- and long-term proportional fairness.
Their performance in realistic complex scenarios will be evaluated
by means of emulation experiments in~\rsec{eval}.
%
%
%
In the remainder of this section we assume, for ease of notation,
that only the destinations with weight $< \infty$ are considered.

\subsubsection{Least-impedance}

The least-impedance (LI) policy is to send the lambda request to
$\bar{d}$, where $\bar{d} = \arg\min_j w_j$.

Quite clearly, this policy is a limit case where there is no
proportional fairness at all: LI is an extremely simple greedy
approach which does not require any state to be maintained at
the e-router.
On the other hand, it may suffer from the herd effect mentioned
above.

\subsubsection{Random-proportional}

The random-proportional (RP) policy is to send the lambda request
to $d$ with probability $1/w_d$.

It is straightforward to see that RP enjoys the property of \textit{long-term
proportional fairness}: over a sufficiently large observation interval
the ratio between the number of lambda requests served by any two
e-computers $i$ and $j$ is $N_i/N_j=w_j/w_i$.
However, in the short-term, fairness is not provided: due to the
probabilistic nature of this policy, there is no mechanism preventing
any destination (or sub-set of destinations) to be selected repeatedly
for an arbitrary number of times, thus making RP arbitrarily unfair
over any finite time horizon.

\subsubsection{Round-robin}

The round-robin (RR) algorithm is inspired from a high-level
discussion in~\cite{Destounis2016} and combines together the
proportional nature of RP with the greediness of LI\@.
With RR we maintain an \textit{active list} of destinations
$\mathcal{A} = \{ w_j | w_j \le 2 \cdot \min_h w_h \}$.
For each destination $j \in \mathcal{A}$ we keep a \textit{deficit
counter} $\Delta_j$ initially set to $0$.
The policy is then to send the lambda request to $d = \arg\min_j
\Delta_j$ and at the same time increase $\Delta_d$ by $w_d$.
Once a destination goes out of the active list, it is
re-admitted for a single ``probe'' after an initial back off time,
doubled every time the probe request fails to meet the
active list admission criterion; when the active set is updated
we decrease all $\Delta_j$ by $\min_h \Delta_h$.
We note that, as a new destination is added, a trivial implementation
following from the algorithm definition would have $\mathcal{O}(n)$
time complexity since all destinations in $\mathcal{A}$ need have
their deficit counter updated.
However, a simple data structure that avoids such linear update can
be adopted: the active destination deficit counters can be stored
in a list sorted by increasing deficit counters, where only the
differences with the previous elements are kept.
For instance, if the deficit counters are $\{4, 6, 7, 7\}$ the data
structure will hold $\{4, 2, 1, 0\}$; since the front element of the
list is always the minimum, the update can be done in $\mathcal{O}(1)$
by simply resetting this value.

\begin{figure}
{\small

  \begin{center}
  \added{
  \fbox{\parbox{3.2in}{

  {\bf initialization}:
  $\mathcal{A} = \mathcal{P} = \emptyset$,
  $\forall j : w_j = e_j = 0, b_j = b_{\min}$
\vspace{0.5em}

{\bf function} {\it leastImpedance()}:

  \hspace{1em} {\bf return} $\arg\min_j \left\{ w_j \right\}$
\vspace{0.5em}

{\bf function} {\it randomProportional()}:

  \hspace{1em} {\bf return} random $d$ with probability $1/w_d$
\vspace{0.5em}

{\bf function} {\it roundRobin()}:

  \hspace{1em} select random $d \in \left\{ j | e_j > now() \land j \notin \mathcal{P} \right\}$

  \hspace{1em} {\bf if} $\exists d$: $\mathcal{P} = \mathcal{P} \cup \{ d \}$

  \hspace{1em} {\bf else}:

  \hspace{2em} $d = \arg\min_j \left\{ \Delta_j | j \in \mathcal{A} \right\}$

  \hspace{2em} $\Delta_d = \Delta_d + w_d$

  \hspace{1em} {\bf return} $d$
\vspace{0.5em}

{\bf function} {\it onReceiveResponse($\delta, d$)}:

($\delta$ is the latency measured for the response from $d$)

  \hspace{1em} {\bf if} {\it leastImpedance} $\lor$ {\it randomProportional}:

  \hspace{2em} $w_d = \alpha \cdot w_d + (1 - \alpha) \cdot \delta$

  \hspace{1em} {\bf else}: {\tt /*} {\it roundRobin} {\tt */}

  \hspace{2em} {\bf if} $d \in \mathcal{P}$:

  \hspace{3em} $\mathcal{P} = \mathcal{P} \setminus \{ d \}$
  
  \hspace{3em} {\bf if} $\delta \leq 2 \cdot \min_j \left\{ w_j | j \in \mathcal{A} \right\}$:

  \hspace{4em} $\forall j \in \mathcal{A}: \Delta_j = \Delta_j - \min_j \{ \Delta_j \}$

  \hspace{4em} $\mathcal{A} = \mathcal{A} \cup \{ d \}$

  \hspace{4em} $b_d = b_{\min}, \Delta_d = w_d = \delta$

  \hspace{3em} {\bf else}:

  \hspace{4em} $b_d = 2 \cdot b_d, e_d = now() + b_d$

  \hspace{2em} {\bf else}:

  \hspace{3em} $w_d = \alpha \cdot w_d + (1 - \alpha) \cdot \delta$

  \hspace{3em} {\bf if} $w_d > 2 \cdot \min_j \left\{ w_j | j \in \mathcal{A} \right\}$: $\mathcal{A} = \mathcal{A} \setminus \{ d \}$

  }}
  }
  \end{center}

}
\vspace{-0.5em}
\caption{Pseudo-code of the algorithms in \rsec{algorithms}.}
\label{fig:pseudocode}
\vspace{-1.5em}
\end{figure}

\added{
The algorithms presented so far in natural language are also described
with the pseudo-code in \rfig{pseudocode} (corner cases not addressed
for simplicity, we refer the interested reader to the real
implementation, available as \textit{open source} for full details),
where $b_j$ is the relative backoff time before a destination is
probed for its inclusion in the active list (starting with minimum
value $b_{\min}$) and $e_j$ is the absolute time when it will be
considered for inclusion in the set $\mathcal{P}$ of probed
destinations.
Note that least-impedance and random-proportional only use $w_j$
state variables.
}

In the following we show that RR achieves both short-term and
long-term proportional fairness.
To this aim, we introduce a more formal definition of the algorithm.
Let $\mathcal{A}$ be the set of active destinations, each with a
deficit counter $\Delta_i$ initialized to $0$.
At time $n$, select destination $x = \arg\min_i\{\Delta_i\}$, breaking
ties arbitrarily;
then, update the deficit counter of the destination selected as follows:
$\Delta_x = \Delta_x + \tilde{\delta_x}$, where $\tilde{\delta_x}$
is an estimate of the time required for the lambda function requested
to complete if forwarded to edge computer $x$.
We denote as $\delta_x$ the \textit{actual} completion time, but
this cannot be known at the time the forwarding decision is taken.
When a new edge computer $y$ is added to $\mathcal{A}$ it is assigned
$\Delta_y = 0$ and all the other deficit counters are updated as
follows:
\begin{gather}
\Delta^{\min} = \min_{i \in \mathcal{A} \setminus y}\left\{\Delta_i\right\} \\
\forall i \in \mathcal{A} \setminus  y : \Delta_i = \Delta_i - \Delta^{\min}
\end{gather}
After each update $\exists i \in \mathcal{A} | \Delta_i = 0$.

\begin{lemma}\label{lemma1}
The difference between any two deficit
counters is bounded by a finite constant equal to the maximum
completion time estimate ($\tilde\delta^{\max}$), i.e.:
\begin{equation}
\max_{i,j}\left\{\Delta_i - \Delta_j\right\} \leq \tilde\delta^{\max}
\end{equation}
\end{lemma}

\begin{proof}
The statement can be proved by contradiction.
Assume:
\begin{equation}\label{eq:lemma1a}
\exists i,j | \Delta_i(n) - \Delta_j(n) = \tilde\delta^{\max} + \epsilon
\end{equation}
Without loss of generality, we can assume that $i$ was the destination
selected at time $n-1$ (it this was not the case, then one can
substitute $n-1$ in the following with the last time $i$ was selected;
meanwhile we can assume that $j$ was never selected otherwise its
deficit counter would have increased, which would have \textit{reduced}
the gap between deficit counters).
In this case, it is:
\begin{equation}\label{eq:lemma1b}
\Delta_i(n-1) = \Delta_j(n-1) = \Delta_j(n)
\end{equation}
Putting together~\req{lemma1a} and~\req{lemma1b} we obtain:
\begin{align}
\Delta_i(n) &= \Delta_j(n) + \tilde\delta^{\max} + \epsilon = \\
            &= \Delta_i(n-1) + \tilde\delta^{\max} + \epsilon
\end{align}
From which it is:
\begin{equation}
\tilde\delta_i = \tilde\delta^{\max} + \epsilon
\end{equation}
Which is impossible for any $\epsilon > 0$ by definition of
$\tilde\delta^{\max}$.
\end{proof}

In the following we assume for simplicity that the completion time
estimates are constant over time, i.e.\ it is $\tilde\delta_i(n) =
\delta_i$ for any destination $i$.
Furthermore, we define $s_i(n)$ as the number of times the destination
$i$ was selected by the forwarding algorithm since $n=0$.

\begin{lemma}[short-term fairness]
For any two destinations
$i,j$ at any time $n$ the difference between their number of services
weighted by the respective completion time is bounded by the maximum
completion time $\delta^{\max} = \max_i\{\delta_i\}$:
\begin{equation}
\forall i,j \in \mathcal{A} : s_i(n) \delta_i - s_j(n) \delta_j \leq \delta^{\max}
\end{equation}
\end{lemma}

\begin{proof}
To prove the statement let us consider the update process of both
$s_i(n)$ and $\Delta_i(n)$:
\begin{equation}\label{eq:lemma2a}
s_i(n) = s_i(n-1) + \begin{cases}
  1 & \text{if } i = \arg\min_j\left\{\Delta_j(n-1)\right\} \\
  0 & \text{otherwise}
\end{cases}
\end{equation}
\begin{equation}\label{eq:lemma2b}
\Delta_i(n) = \Delta_i(n-1) + \begin{cases}
  \delta_i & \text{if } i = \arg\min_j\left\{\Delta_j(n-1)\right\} \\
  0 & \text{otherwise}
\end{cases}
\end{equation}
From~\req{lemma1a} and~\req{lemma1b} it follows that:
\begin{equation}
\Delta_i(n) = s_i(n) \delta_i
\end{equation}
which proves the statement with Lemma~\ref{lemma1}.
\end{proof}

\begin{table*}[tbp]%
\caption{V}%
\centering%
\include{tables/}%
\label{tab:}%
\end{table*}%
C
{schedexample}%
{Visual forwarding example used in Lemma~\ref{lemma3} with three
destinations with $\delta_1 = 2$, $\delta_2 = 3$, $\delta_3 = 4$.}

\begin{lemma}[long-term fairness]\label{lemma3}
Over a long enough
time horizon the number of services of any destination $i$ is
inversely proportional to its completion time $\delta_i$, i.e.:
\begin{equation}
\exists i,j \in \mathcal{A}, n >0 | \frac{s_i(n)}{s_j(n)} = \frac{\delta_j}{\delta_i}
\end{equation}
\end{lemma}

\begin{proof}
Let us start with the assumption that $\delta_i \in \mathbb{N}$.
In this case, using the visual example in~\rtab{schedexample},
it is easy to see that there is a time $n^*$ where $\Delta_i =
\Delta_j$ for any $i,j$, which can be computed as:
\begin{equation}
n^* = \sum_i \frac{T}{\delta_i}
\end{equation}
where $T = lcm_i\{\delta_i\}$, which gives:
\begin{equation}
s_i(n^*) = \frac{T}{\delta_i}
\end{equation}
from which the lemma follows.
The statement can be easily extended to the case of $\delta_i \in
\mathbb{Q}$ as follows.
Find $D$ as the LCM of the denominators of $\delta_i$, then set
$\delta'_i = D/den(\delta_i)$.
We now have new $\delta'_i$ values that are in $\mathbb{N}$ by
construction and the same method as above can be used.
\end{proof}

%
%
%

%% file: scalability.tex
In this section we study the scalability of our proposed architecture
in terms of the number of serverless functions offered by the
e-computers.

In~\rsec{contribution} we foresee a flat overlay where
each e-router is notified of the existence of \textit{any} lambda function
offered by an e-computer by the e-controller, including a means to
reach it (such a \ac{URI} or a service end-point).
This means that every e-router is required to keep one entry in its
e-table for every lambda function in the network (say $\Lambda$).
In~\rsec{algorithms} we have put forward algorithms that are simple
and efficient to both keep the entry's weight updated and select the
destination for any incoming lambda, yet performance may suffer as
$\Lambda$ becomes very big, especially if the e-routers are low-end
devices such as \acp{SBC}.

To verify this claim, we have carried out the following experiment
on a Raspberry Pi~3 Model~B (RPi3), which is ``representative of a
broad family of smart devices and appliances''~\cite{Morabito2017a},
where we have installed an e-router and added an artificial number
of entries in its e-table.
Then, we have started a number of clients repeatedly asking for
a lambda function to be executed.
In particular, we have used 10 clients, as the minimum number for
which the computation resources of the RPi3 were saturated.
The e-routers have been modified for the purpose of the experiment
so that, instead of actually forwarding the lambda request to an
e-computer and reporting back the result to the client, it just
replied with a successful but empty response for any lambda.
All other phases, such as weight computation, destination
selection, and any housekeeping operations, were performed exactly
as in a real environment.
\textit{This experiment provides us with an upper bound of the number of
lambda functions that can be processed concurrently by an e-router
on a RPi3 as the e-table size increases.}
We have run 10 experiments but the variance was negligible compared
to the mean values, thus we report only the latter in the plot;
similarly, we report only the results with RR as the destination
selection algorithm, because the other algorithms gave similar
results leading to the same conclusions.



\myfigeps%
{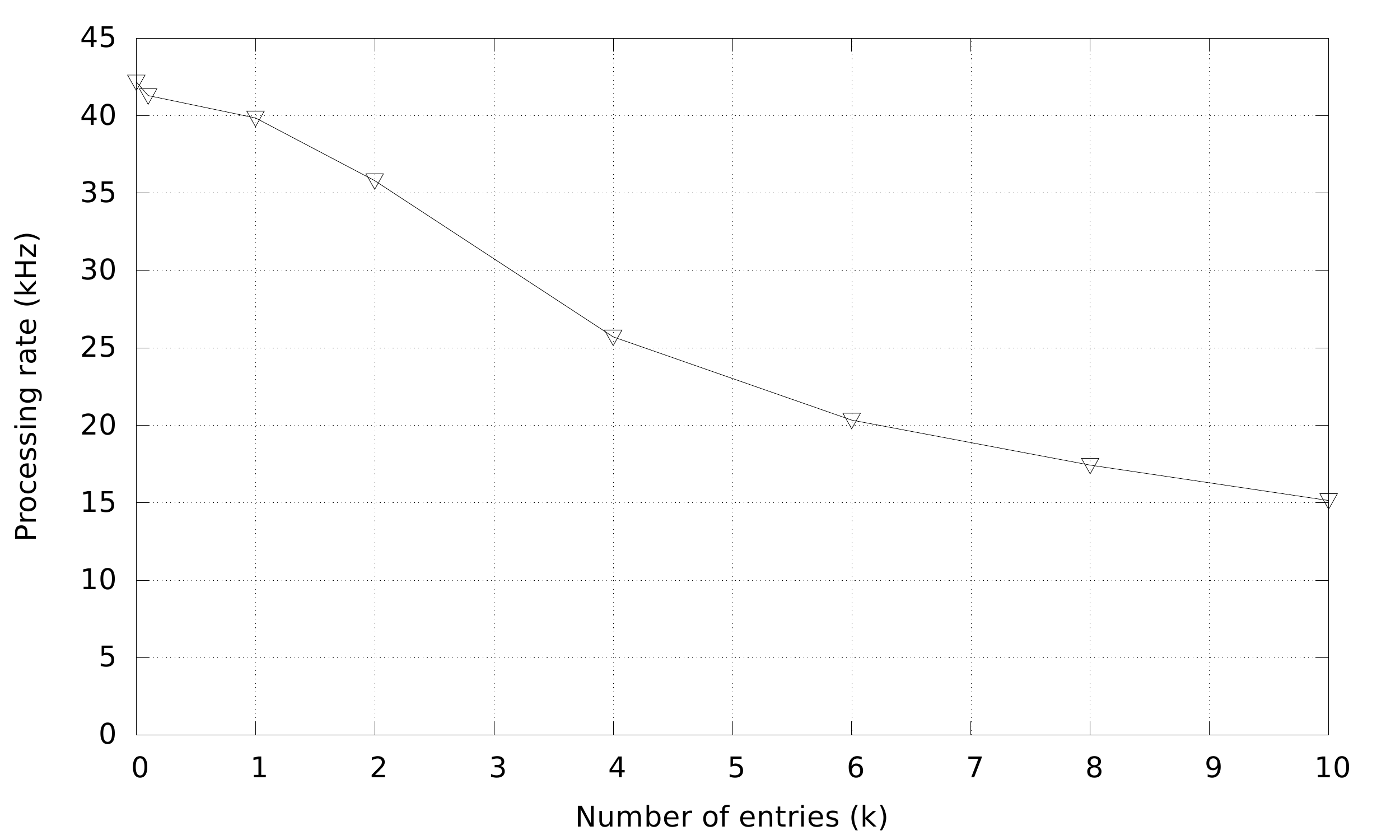}%
{RPi3 e-router processing rate vs.\ num.\ forwarding table entries.}

%
In \rfig{rpi3-single} we show the normalized
processing rate, which decreases steadily as the e-table increases at steps
of 1,000 entries at a time, showing an asymptotic behavior with
very big tables that is consistent with a $\mathcal{O}(\log(\Lambda))$
average time complexity, which can be inferred \textit{a priori}
since the algorithms involved require a look-up in a sorted data
structure to both retrieve the weight and select the next destination.
%
%
The results confirm the intuition that the size of the e-table may
become a choke point in a production environment as the number of
lambda functions increases to realistic values, at least with devices
with limited processing capabilities.

\myfigeps%
{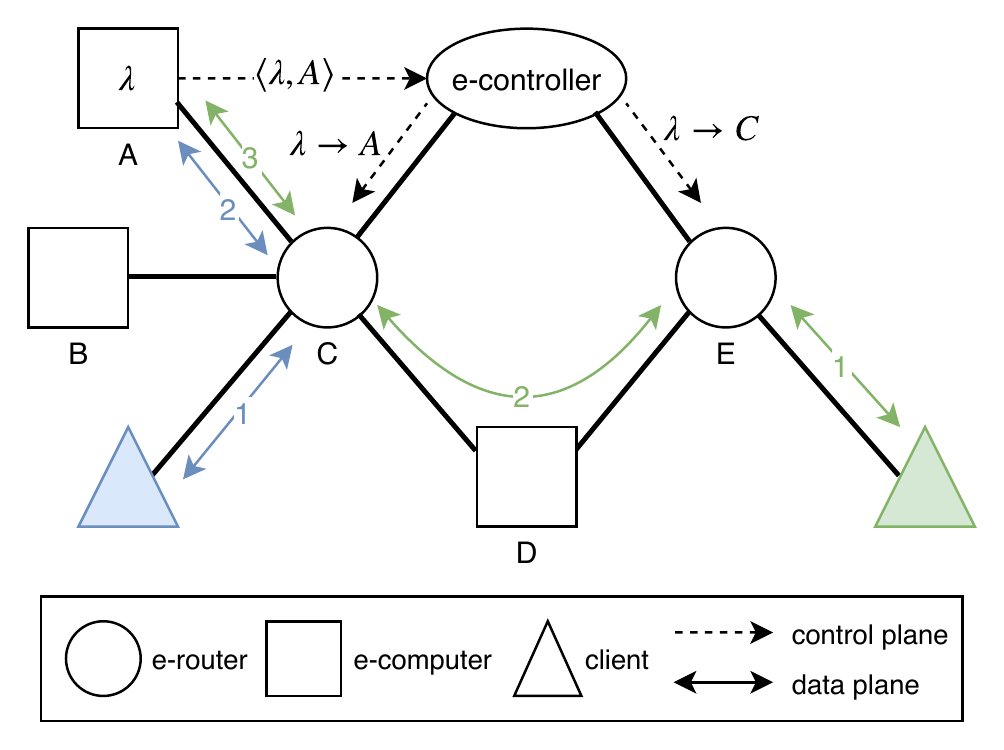}%
{Extension of the proposed architecture to improve scalability.}

To enhance scalability while keeping the same architecture and
functions of the e-routers and e-computers, we propose the following
optional extension of the e-controller.
Recall that the e-controller is notified of the lifecycle of
lambda functions on computers, i.e., start-up and tear-down of
their serverless executors.
Rather than announcing a new lambda function from the e-computer
to all the e-routers, the e-controller can announce that
to only \textit{some} e-routers, while the others are notified that
the lambda can be executed by one of the latter.
Let us explain this concept with the example in~\rfig{scalability-arch},
which shows what happens as the e-computer A announces $\lambda$
to the e-controller.
The latter announces to C that $\lambda$ can be executed by A, but it
tells E that $\lambda$ can be executed by C.
Obviously, with a single instance of $\lambda$ there is no advantage
since both e-routers will have exactly one entry in their e-table.
However, imagine that a new lambda is then started on e-computer B
and the e-controller does the same as before: in this case, the
e-table of C grows by one entry, but the e-table of E remains
the same because it already knows that it is possible to execute
$\lambda$ via C.
In the example in~\rfig{scalability-arch} we also show the
data plane flows associated to two clients, connecting to the
edge computing domain through C and E, respectively.
Therefore, this \textbf{hierarchical overlay} allows the e-tables
to grow less than linearly with the number of lambda services.
In the supplementary material we also explain how to deal with
forwarding loops, which may arise in more complex network topologies.
The solution is to distinguish between \textit{final} and
\textit{intermediate} destinations when advertising e-table entries,
the former being preferred over the latter when forwarding.

Such a two-tier overlay could be easily extended to the case of $n$
tiers, however we do not consider this case because we believe that
just one indirection level is sufficient to achieve scalability
while keeping the protocol and processing overhead to a bare minimum.
In \rsec{scalability:algo} below we propose an algorithm that can
be used by the e-controller to determine which entries to announce
to the e-routers, based on external knowledge of the network topology.

\myssec{An e-table distribution algorithm}{scalability:algo}

We now propose a practical algorithm that can be used by the
e-controller to realize the hierarchical overlay described above.
First, we note that it is a responsibility of the e-controller
alone to announce entries in such a way that clients connecting
from \textit{any} e-router can always consume all the available
services.
This basic functional requirement translates into: for every
lambda, any e-router must have in its e-table either a final
destination to an e-computer offering that lambda, or
intermediate destinations which, in turn, have at least one
such final destination.
Otherwise, a lambda request might get ``stuck'' at an e-router.
Second, we argue that the e-controller should use network topology
information to take its decisions:
\begin{enumerate*}[label=(\roman*)]
  \item such information is readily available from the \ac{SDN}
  controller and it is likely to change slowly over time, at least
  as far as the e-routers and e-computers are concerned\footnote{%
  Remember that we assume that allocation of lambdas to e-computers
  requires instantiations of \acp{VM}/containers, thus it is adjusted
  on a longer time frame with respect to the dynamics of weight
  updating and lambda request forwarding considered in this work.};

  \item disregarding connectivity may lead to inefficient
  allocation of e-table entries, e.g., a final entry for an
  e-computer may be installed on an e-router whose network cost
  is very high, thus forcing all other e-routers going through
  that to consume much more network resources than needed, in
  addition to adding significant delay to the data plane.
\end{enumerate*}

Therefore, in the following we assume that the e-controller is aware
of the distance matrix $\{ d_{ij} \}$ between any two edge nodes
$i$ (source) and $j$ (destination), which in a real environment can
be acquired through the \ac{SDN} controller.
We further simplify the problem by assuming that the e-controller
identifies for each e-computer a \textit{home} e-router.
For all the lambdas of an e-computer, the home e-router is always
advertised as final to itself and as intermediate to every other
e-router.
We believe this approach is very practical since it allows the
e-controller to pre-compute for all possible e-computers their
respective home e-routers, saved in a dedicated look-up table
used whenever a new lambda appears.
This is very useful since the rate at which the topology changes
is expected to be much lower than that of lambda functions' lifecycle
changes, which may occur based on some autoscaling feature available
in the edge computing domain, as mentioned earlier.

Under this assumption, the operation of the e-controller becomes
straightforward:
\begin{itemize}[leftmargin=*,label={--}]
  \item as a new lambda function is announced from e-computer $c$:
  look up $c$'s home router $\bar{r}$; announce a final destination
  entry towards $c$ to $\bar{r}$; announce an intermediate
  destination towards $\bar{r}$ to any $r \ne \bar{r}$, unless already
  announced;

  \item as a lambda function is torn down by e-computer $c$:
  look up $c$'s home router $\bar{r}$; remove the final destination
  entry towards $c$ from $\bar{r}$; remove an intermediate
  destination towards $\bar{r}$ from any $r \ne \bar{r}$, unless
  there is at least one other final destination towards $c' \ne c$
  in $\bar{r}$ for the same lambda.
\end{itemize}

Thus, the only remaining issue is \textit{how to select the home e-router of
an e-computer.}
To this aim, we propose to pursue one of the following two objectives:
\textit{min-max}, where the choice of the home e-router minimizes
the maximum cost that has to be paid by a client accessing from any
e-router to reach the home e-router and from it the final destination
e-computer; or, \textit{min-avg}, where we strive to minimize the
same cost paid by clients accessing from all e-routers, on average.
In any of the two policies we break ties based on the result that
would be obtained if using the other policy (further ties are
broken arbitrarily).
Both policies are captured by the following objective (we use
the sum instead of the average since the number of e-routers is a
constant):
\begin{align}
\label{eq:scalability:objective}
& \min_{i \in \mathcal{R}} \left\{ \Omega \max_{j \in \mathcal{R}} \left( d_{ic} + d_{ij} \right)  +
  \omega \sum_{i \in \mathcal{R}} \left( d_{ic} + d_{ij} \right) \right\} = \nonumber \\
& \min_{i \in \mathcal{R}} \left\{ \Omega \left( d_{ic} + \max_{j \in \mathcal{R} d_{ij}} \right) +
  \omega \left( \left| \mathcal{R} \right| d_{ic} + \sum_{i \in \mathcal{R}} d_{ij} \right) \right\},
\end{align}
where $\mathcal{R}$ is the set of all e-routers, and $\Omega$, $\omega$
are constant factors that can be used to shift from min-max to
min-avg: if $\Omega \gg \omega$ then the policy is min-max, otherwise
if $\Omega \ll \omega$ it is min-avg, where the ratio between the
two must be large enough that the maximum always overpowers the
average, e.g., $\Omega/\omega > 2 \left| \mathcal{R} \right| D$,
where $D$ is the connectivity graph's diameter.

Regardless of the objective function, the search of the home e-router
of a single e-computer is $\mathcal{O}(|\mathcal{R}|^2)$, hence the
overall process to find the home e-routers for all e-computers if
$\mathcal{O}(|\mathcal{R}|^2 |\mathcal{C}|)$, if $\mathcal{C}$ is
the set of the e-computers.
%
%
The algorithm must be executed only by the e-controller and
only upon changes of the connectivity of edge nodes or the
set of lambda functions offered by them.
We believe that in many use cases of practical interest both these events
will be much less frequent than lambda function execution, i.e.,
order of magnitude of minutes and above for the former vs.\ seconds
and below for the latter.
A more optimized algorithm, tailored to the specific deployment,
can be devised and implemented instead on the e-controller, if
needed, with no impact at all on the other system components.

\myfigeps[width=2in]{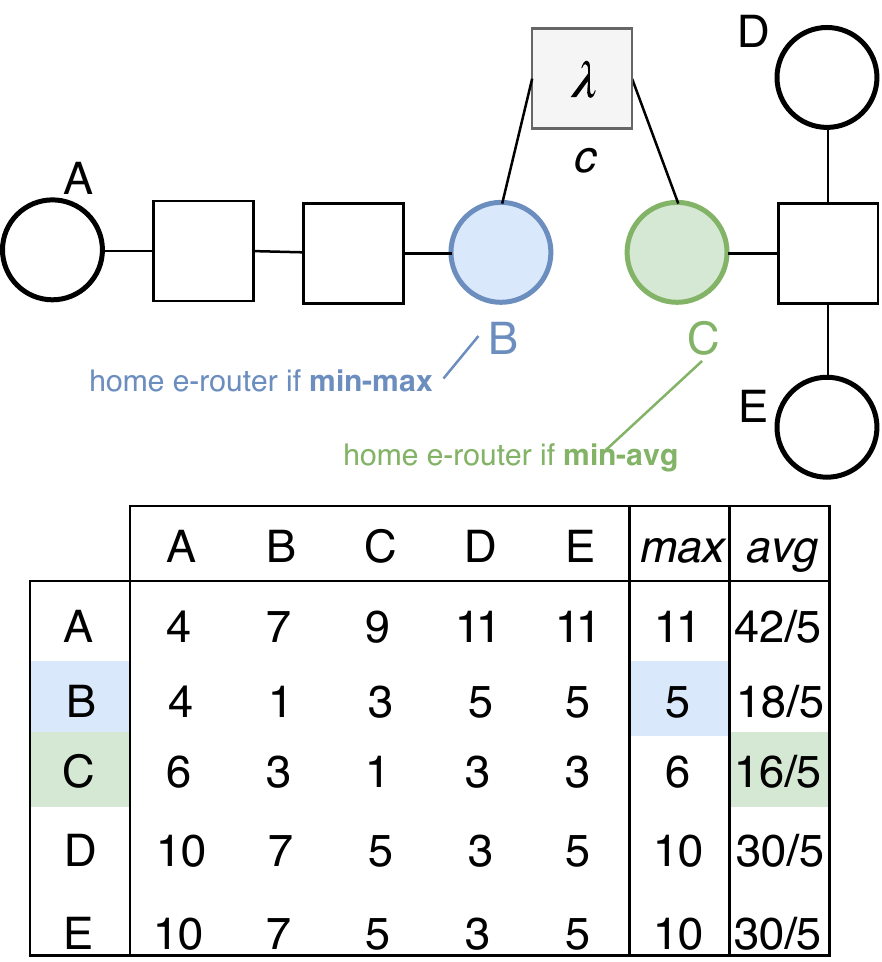}{%
  Determining the home e-router in an example topology: min-max vs.\
  min-avg.}

In~\rfig{scalability-example} we show an example of
determining the home e-router of an e-computer offering $\lambda$.
We assume that the cost of all hops is equal to 1 and that the
distance matrix is symmetric (note that this is just for the
sake of the example: the algorithm in this section supports
asymmetric paths).
On the right we report a matrix where element $x_{ij}$ is the sum
of the cost to go from e-router $i$ to the e-router $j$ and from
there to the e-computer $c$, i.e.  $d_{ic} + d_{ij}$ as
in~\req{scalability:objective}.
In the example, B is selected as the home e-router if min-max is
used because the maximum path  D---B---$c$ (or E---B---$c$, both
with cost 5) is smaller than the maximum path that would exist if
any other e-router is selected as home, including C for which the
maximum path would be A---C---$c$ (cost 6).
However the e-router C is more central to the network topology, as
reflected by the smaller average distance, and it is therefore
selected as home of $c$ with min-avg.

We note that the proposed approach may lead, in general, to sub-optimal
routing of lambda requests/responses.
For instance, with reference to~\rfig{scalability-example} with a
min-avg policy, the path followed by a lambda request from A in
terms of ``forwarding hops'' is A---C---$c$, but in ``networking
hops'' this requires A---B---$c$---C---$c$, which is clearly
inefficient since, in principle, the request could have stopped at
$c$ the first time it got there.
However, we believe the ends justify the means: by tolerating
sub-optimal transfer of lambda requests, we can keep the design of
the solution, based on the concept of ``home e-router'', simple and
efficient.
Anyway, in many cases of practical interest the performance degradation
due to the detours could be negligible, as shown by the results of
the experiments in~\rsec{eval:large} obtained using a realistic
network map.

\myssec{Analysis}{scalability:analysis}

In qualitative terms, the hierarchical overlay proposed above trades
off \textit{response delay} (increased because of one additional
hop from the client to the e-computer) for \textit{computational
complexity} in the e-routers (decreased by having smaller e-tables,
which occupy less memory and yield a faster look-up).
In this section we perform a quantitative analysis of these two
aspects in simplifying conditions, which will be complemented
in~\rsec{eval} by emulation experiments in more realistic environments.
%

We start with the computational complexity.
%
%
The actual size of an e-table will, by necessity, depend on the
relative distance of the e-routers and the e-computers offering each
lambda, and it cannot be predicted \textit{a priori}.
Rather, it is easy to forge limit cases where the maximum size of
the e-table is not reduced at all by the hierarchical scheme proposed.
Consider for instance a network where all e-computers have to pass
through an e-router, acting as a sort of default gateway, which
then connects to all the other e-routers in the edge domain: in
this case the e-table of the ``gateway'' will contain all the lambda
in the network (as it would if a flat overlay was used), while all
the others would contain a single entry pointing to the ``gateway''
itself.
Such degenerate cases are of limited interest, since they would
call for \textit{ad hoc} solution anyway.

Thus, we argue that a case where the e-routers and e-computers are
distributed rather uniformly over the edge domain, in a topology
sense, is much more realistic, especially since we consider \ac{IoT}
environments whose nature is often distributed over large coverage
areas.
Under these reasonable assumptions, every e-router is bound to
have approximately the same chances as any other to be selected
as the home e-router for a given e-computer/lambda.
Thus, reusing the same notation as above, an approximation of the
average number of entries in an e-table for every $\lambda$ served
by $\mathcal{C}_\lambda$ e-computers is:
\begin{equation}
  \label{eq:num-entries}
n_{entries} \approx
\begin{cases}
  |\mathcal{C}_\lambda| & \text{if } |\mathcal{C}_\lambda| \ll |\mathcal{R}| \\
  |\mathcal{R}| + \frac{|\mathcal{C}_\lambda|}{\mathcal{R}} & \text{if } |\mathcal{C}_\lambda| \gg |\mathcal{R}| \\
\end{cases}.
\end{equation}
An explanation of~\req{num-entries} is the following: when the
number of e-computers serving a given lambda is small, compared to
the number of e-routers, then there are little or no advantages in
using a hierarchical overlay, since in a uniformly distributed
topology the chance of the few lambdas ``colliding'' in the same
e-router are small anyway; however, if this is the case, then it
is not \textit{necessary} to use a hierarchical overlay, because
its only motivation is to limit the size of e-tables as the number
of lambda functions increases significantly.
In this case, i.e., the bottom branch of~\req{num-entries}, (almost)
all the e-routers will have at least one e-computer offering a given
lambda, hence every e-router will have to have one entry for
each $r \in \mathcal{R}$, in addition to the lambda functions for
which it is a home router, which will split evenly among all the
e-routers in the network, hence amounting to
$\frac{|\mathcal{C}_\lambda|}{\mathcal{R}}$.
In a large edge computing domain, where scalability must be considered
as as potential issue, we definitely expect to have many e-routers
($|\mathcal{R}|$ is big), hence a two-tier overlay is sufficient
to keep the pace at which e-table sizes grow much slower than the
lambda increase rate.

\myfigeps%
{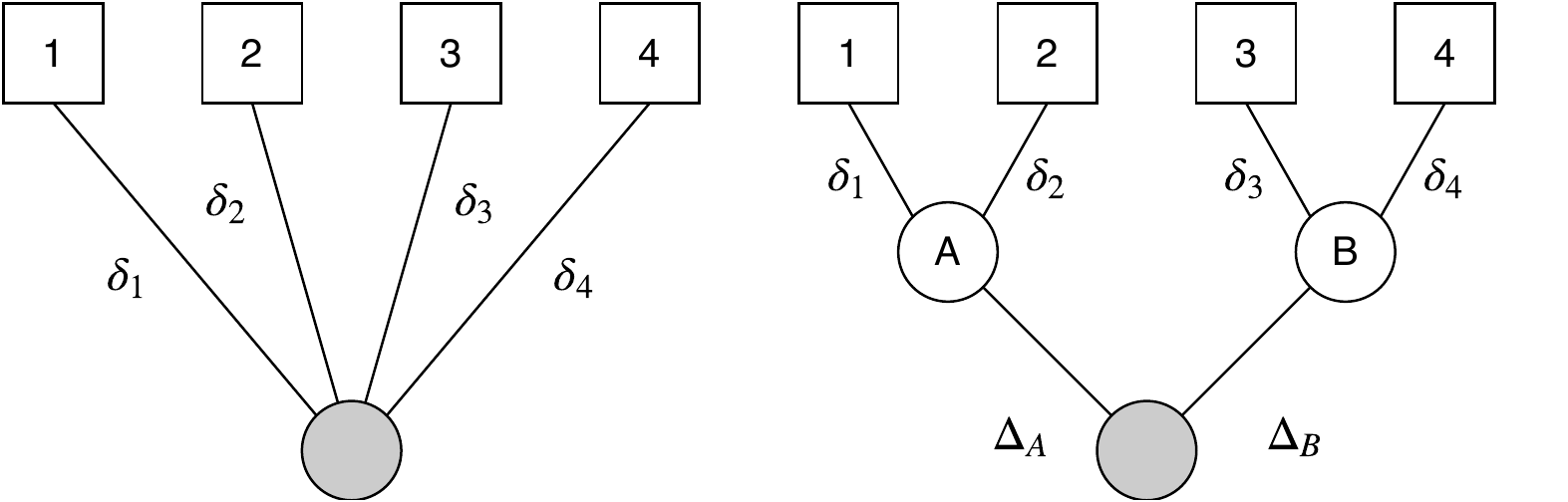}%
{Example with four e-computers to illustrate the effect of using a
hierarchical overlay on the average delay.}


We now focus on the increased response delay.
%
To this purpose, let us consider the example
in~\rfig{average_delay_example}, showing on the left-hand side an
e-router having on its e-table four final destinations, numbered
from 1 to 4.
Under the notion of proportional fairness introduced
in~\rsec{algorithms:destination}, assuming for simplicity that the
delays remain constant over time and over a sufficiently long period,
every destination $i$ will be selected more frequently proportionally
to the inverse of its delay $\delta_i$.
This means that the average delay experienced, in general with $n$
destinations, is:
\begin{equation}
  E[\delta] = \frac{\sum_{i=1}^n \delta_i \frac{1}{\delta_i}}
  {\sum_{i=1}^n \frac{1}{\delta_i}} =
  \frac{n}{\sum_{i=1}^n \frac{1}{\delta_i}},
\end{equation}
which is the harmonic mean of the delays, which in the example is:
\begin{equation}
  \label{eq:delay-flat}
  E[\delta] = 4
  \frac{\prod_{i=1}^4 \delta_i}
  {\sum_{i=1}^4 \frac{1}{\delta_i} \prod_{i=1}^4 \delta_i}
\end{equation}
Let us now see what happens if another level of forwarding is added,
by introducing e-routers A (home to e-computers 1 and 2) and B (home
to e-computers 3 and 4), assuming that reaching A (B) incurs an
additional delay $\Delta_A$ ($\Delta_B$).
By substitution and simple algebraic simplification, we obtain:
\begin{equation}
  \label{eq:delay-overlay}
  E[\delta] = 4
  \frac{\prod_{i=1}^4 \delta_i + G_A^2 \Delta_B A_B + G_B^2 \Delta_A A_A + \Delta_A \Delta_B A_A A_B}
  {\sum_{i=1}^4 \frac{1}{\delta_i} \prod_{i=1}^4 \delta_i + 2 (\Delta_A + \Delta_B) A_A A_B},
\end{equation}
where $G_A = \sqrt{\delta_1 \delta_2}$ is the geometric mean of the
delays incurred when reaching the lambda through A and $A_A =
\frac{\delta_1 + \delta_2}{2}$ is their arithmetic mean (and likewise
for $G_B$ and $A_B$).
We note two properties from~\req{delay-overlay}.
Firstly, if the additional forwarding delays $\Delta_x$ are small
compared to the processing delays $\delta_y$ then~\req{delay-overlay}
becomes exactly~\req{delay-flat}, because all the addends after the
first in both the numerator and the denominator become negligible
with respect to the first component.
We expect this assumption to be true in many cases of practical
interest, since it basically means that the latency due to transferring
the input and output of a lambda is much smaller than the time it
takes an e-computer to process it.
Secondly, even when this is not applicable, it is clear
from~\req{delay-overlay} that the variations to the average delay
introduced by the $\Delta_x$ are smooth, i.e., they do not change
abruptly the behavior dictated by the $\delta_y$ values.
%

%% file: eval.tex
In this section we provide a comprehensive experimental performance analysis
of the contributions proposed in this work by means of three different
scenarios, designed to validate in particular one aspect of the
overall solution proposed.
Specifically, the decentralized architecture (\rsec{contribution})
is studied in a scenario with a grid network topology (\rsec{eval:grid}),
also illustrating the behavior of the destination selection algorithms
in \rsec{algorithms:destination}; the latter are also subject to
analysis in the second scenario (\rsec{eval:tree-wsk}), which however
focuses on showing the benefits of explicit congestion notification
as proposed in \rsec{algorithms:weight}.
The results will show that our proposed framework, with multiple
decentralized serverless platforms, can achieve almost the same
performance as a \textit{hypothetical} \added{distributed} system running
at the edge with same aggregate capacity.
Finally, the e-router scalability described in \rsec{scalability}
is assessed in a large-scale realistic scenario (\rsec{eval:large}),
showing a reduction of the size of e-tables of about 20\%, compared
to a flat overlay, without a noticeable degradation in terms of the
delay (which in some cases is even decreased, as explained later).
Further results are provided as supplementary material.

\myssec{Environment and methodology}{eval:environment}

We have used our own performance evaluation framework with
real applications running in Linux namespaces interconnected via a
network emulated with Mininet\footnote{\url{http://mininet.org/}}.
The interested reader may find full details in~\cite{Cicconetti2019a},
which also describes the details of the implementation of the
e-routers and e-controller.
As executors we have used both OpenWhisk (as in~\cite{Baresi2019})
running in Docker containers\footnote{\url{https://www.docker.com/}}
and emulated e-computers, which provide responses to lambda requests
based on the simulation of processing of tasks in a multi-processing
system, also illustrated in~\cite{Cicconetti2019a}.

%
%
%
The experiments have been run on a Linux Intel Xeon dual socket
workstation that was not used by any concurrent demanding application.
%
%
The smoothing factor $\alpha$ in \req{weightupdate} was set to 0.95
based on preliminary calibration experiments whose results are not
reported here.
%
%
In the plots we report confidence intervals (unless negligible)
with a 95\% confidence level, computed over 20-30 independent
replications of each experiment, depending on the scenario.





\myssec{Grid scenario}{eval:grid}

\myfigeps{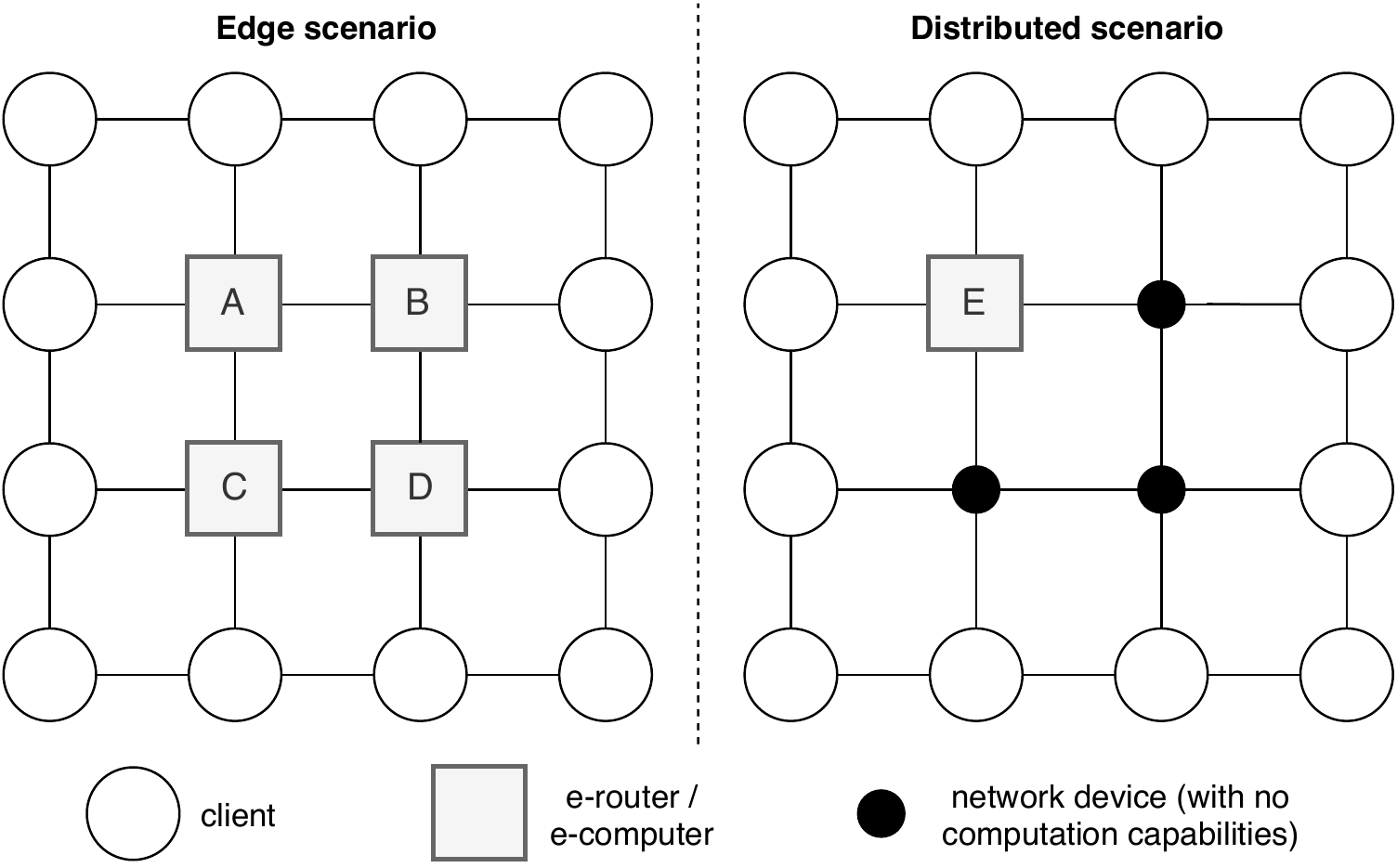}{%
Grid scenario network topology (\rsec{eval:grid}).}

We begin our analysis with a scenario in a grid topology, shown in
\rfig{grid-topo} (left part), with 100~Mb/s bandwidth / 100~$\mu$s
latency links.
As in~\cite{Destounis2016}, the executors are in the middle of the grid,
while the exterior nodes act as clients; every host with an executor also
hosts an e-computer.
We compare our proposed solution (called \textit{Edge/dynamic},
using RR as a destination algorithm) with an alternative with no
e-routers, where clients simply request execution of lambda functions
from the closest executor (called \textit{Edge/static}).
We also run experiments in a slightly different
setup, with a single more powerful executor having computational
capabilities equal to the sum of all the executors in the other
setup.
Such an environment, illustrated in the right part of \rfig{grid-topo}
and called \textit{\added{Distributed}}, is representative of a traditional
serverless deployment (like in the left part of \rfig{distr-vs-edge}).
The results obtained with \added{Distributed}, therefore, are intended only
as a reference, not in comparison to those in Edge environments.

In the first batch of results, the executors use OpenWhisk: nodes
A--D are reserved two CPU cores each, while node E is given eight.
We have implemented two toy lambda functions, which stress respectively
I/O and CPU.
Every client requests their execution with an \ac{IPP} pattern with
busy / idle periods equal to 2~s / 6~s, on average; within a busy
period, the client triggers 3 lambda functions per second;
lambda functions that would extend the busy period because of
accumulated delay are dropped, instead.
Clients are independent of one another.
We ran experiments with a variable number of clients.

\myfigeps{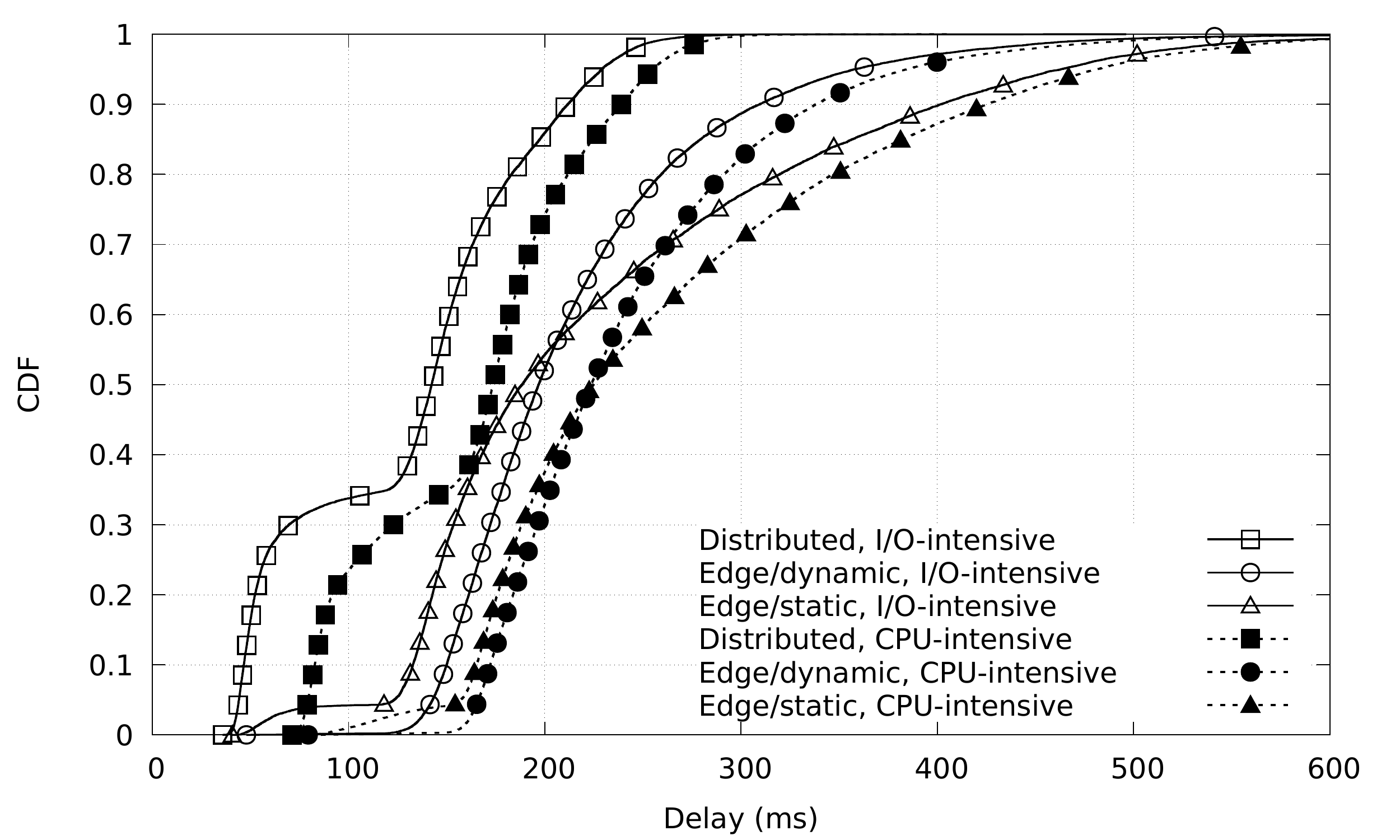}{%
Grid scenario (with OpenWhisk executors):
CDF of delay with I/O- and CPU-intensive tasks, with 24 clients.}

We first evaluate the delay, defined as the time between when the
client is scheduled a lambda function and the time when it receives
a response.
In \rfig{grid-wsk-delay} we show the CDF of the delay with 24
clients, corresponding to a moderate overall load.
With both types of lambda functions, using e-routers is beneficial
in terms of high quantiles of the delay: a static allocation provides
similar (or better) performance in those times when there are
fewer requests to a given executor, but when the load increases
it is unable to balance in the pool of resources available.
The delay with \added{Distributed} is smaller than with Edge: this confirms
the intuition that serverless in a cloud-like environment is easiest
with mainstream technologies, though not always possible because
of deployment constraints.

\myfigeps{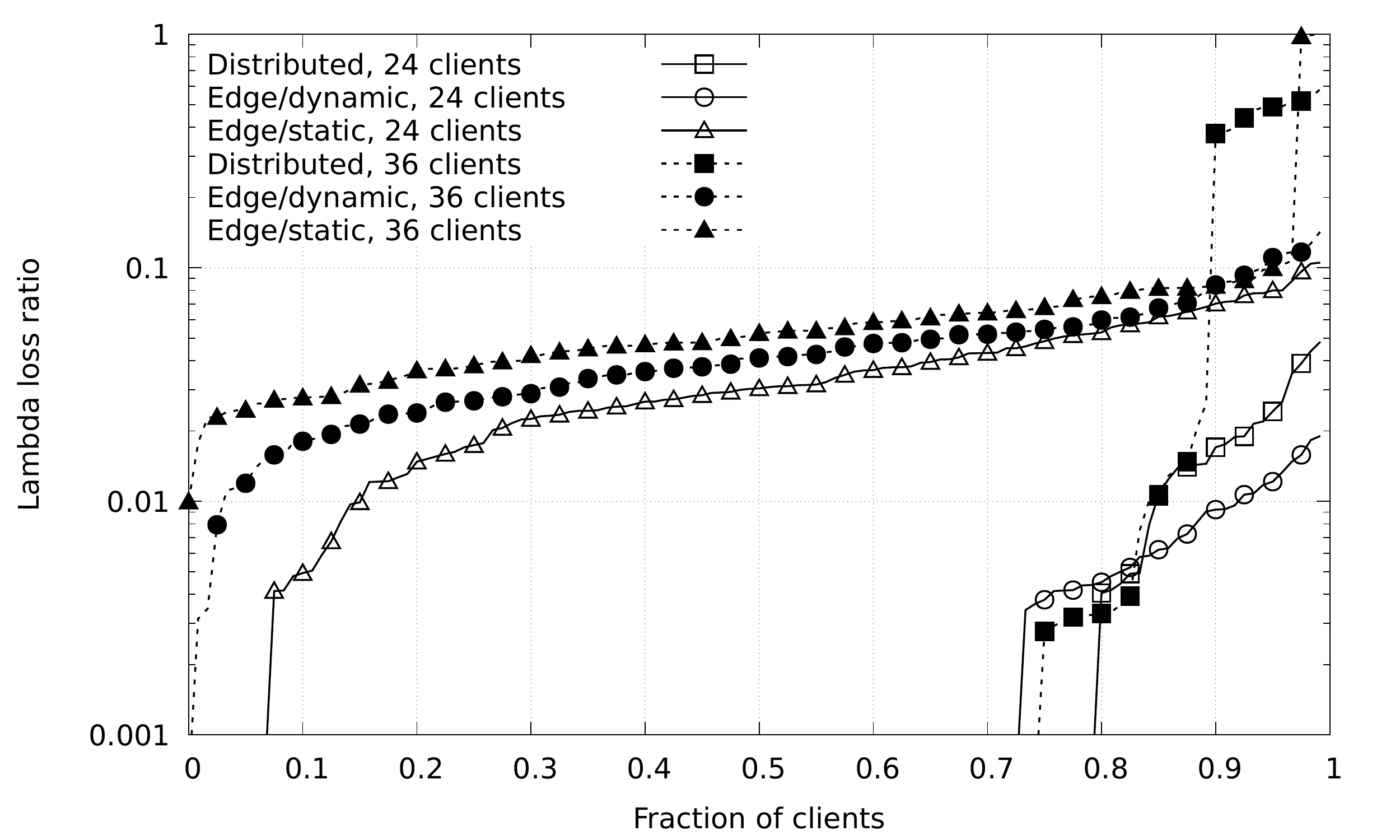}{%
Grid scenario (with OpenWhisk executors):
Lambda loss ratio with CPU-intensive tasks, with 24 and 36 clients.}

In \rfig{grid-wsk-loss} we show the lambda loss ratio, defined as
the ratio of lambda function calls that the client refrains from
executing to avoid overrunning the active periods over the total
scheduled.
We report only the results with CPU-intensive tasks (those with
I/O-intensive tasks are similar and omitted for lack of space),
with 24 and 36 clients.
As can be seen, with both loads a static allocation exhibits poor
performance, with most clients experiencing a non-negligible loss.
It is interesting to note that with 24 clients the Edge/dynamic and
\added{Distributed} curves cross a little above 80\% of clients: while there
are more clients with non-negligible loss in the former, the clients
experiencing some loss with \added{Distributed} have a higher loss than
with Edge/dynamic.
This is an effect of the fairness property of the RR algorithm
illustrated in \rsec{algorithms:destination}, which is confirmed
with 36 clients: Edge/dynamic is the only curve that does not rise
steeply as the fraction of clients in the $x$-axis increases.

\myfigeps{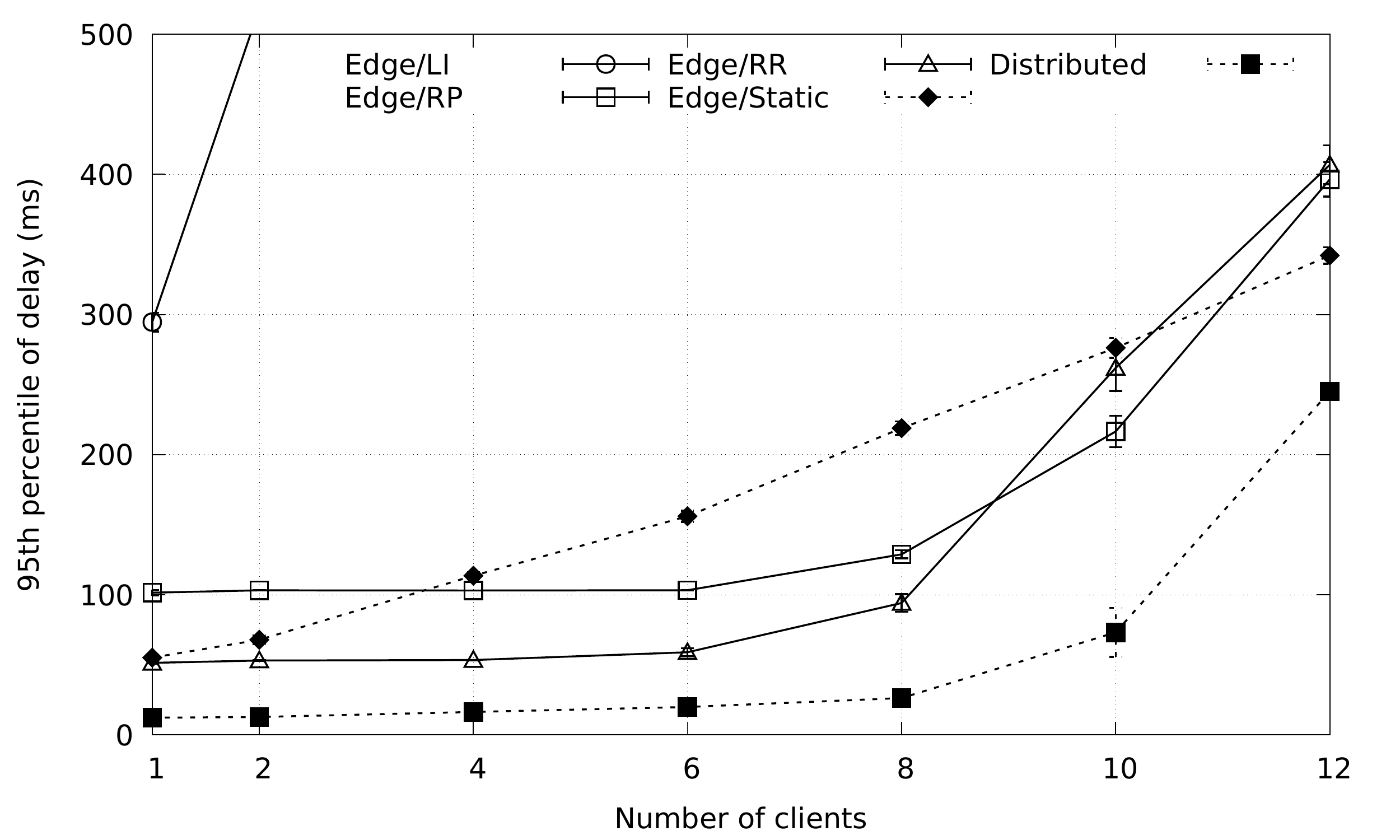}{%
Grid scenario (with emulated e-computers):
95th percentile of delay vs.\ number of clients.}

In the following, we report results obtained by substituting the
OpenWhisk servers with emulated e-computers, which allowed us to
analyze the case with heterogeneous CPUs.
In fact, we set the (simulated) speed of the CPU of executors A/B/C/D
respectively as 1,000/2,000/3,000/4,000~MIPS, whereas E in the
\added{Distributed} case has the sum of the speed of A--D, i.e., 10,000~MIPS.
We increase the number of clients from 1 to 12; results with all
the algorithms in \rsec{algorithms:destination} are reported.
In \rfig{grid-emul-out-095} we show the 95th percentile of delay
as the number of clients increases.
As expected, the \added{Distributed} curve lies below all the others.
Furthermore, we note that Edge/LI performs very bad even at low loads:
this is because of the ``herd'' effect, which makes hot spots
the executors with fastest CPU.
A static allocation performs worse than Edge/RR and Edge/RP,
except at very low loads.
Finally, Edge/RR exhibits a smaller 95th percentile of delay than
Edge/RP, but only until the system becomes overloaded (i.e., with
8 clients or less): after that, the performance of RP is comparable
or better than that of RR.

\myfigeps{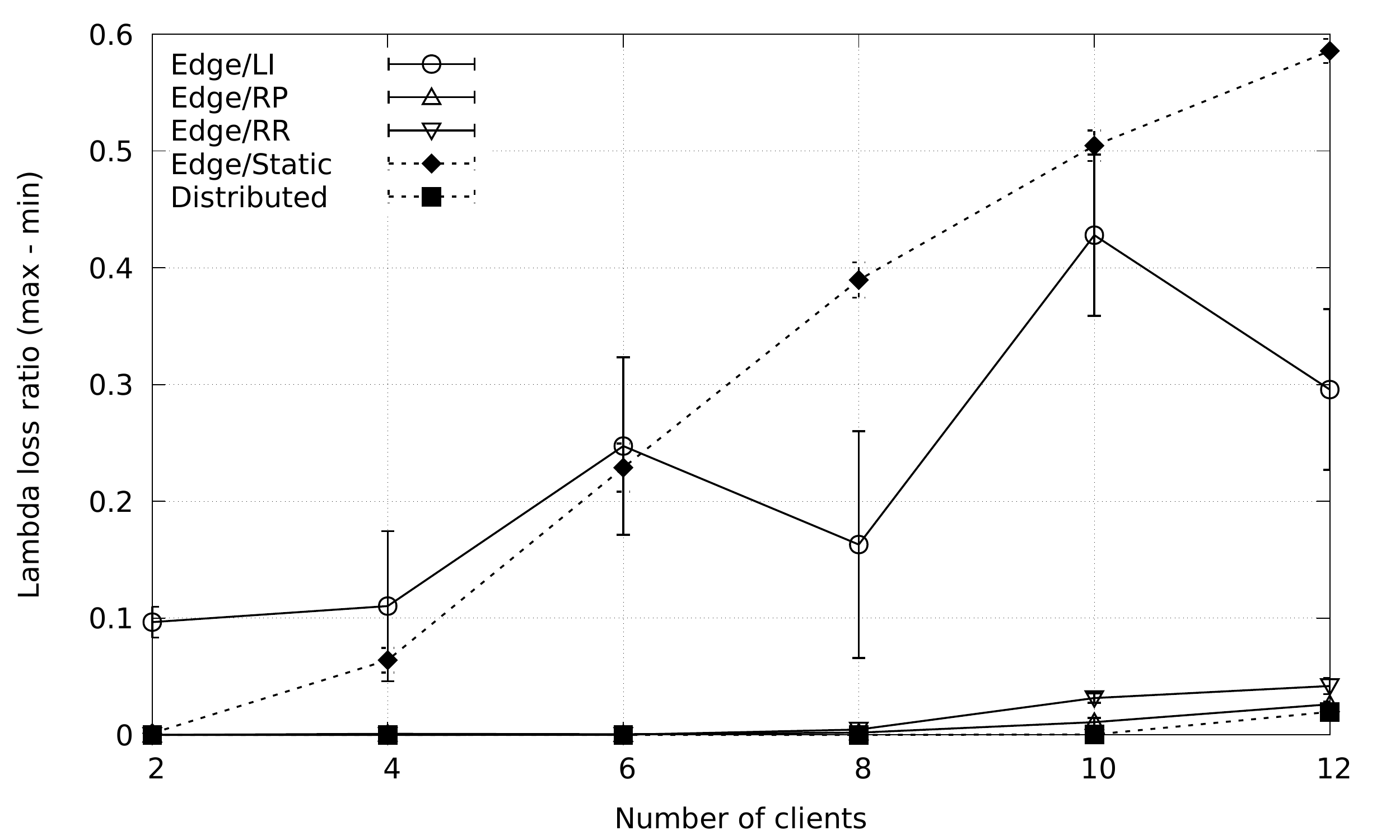}{%
Grid scenario (with emulated e-computers): Difference between the
maximum and minimum lambda loss ratio vs.\ number of clients.}

In \rfig{grid-emul-loss-fair}, we complement the results above with
a measure of the fairness, as the difference between the maximum
and minimum lambda loss ratio among all the clients.
The behavior of Edge/LI is extremely erratic since it tries to
always direct the clients to the fastest executors.
On the other hand, Edge/RR and Edge/RP exhibit excellent fairness,
thus confirming on the field the theoretical analysis in
\rsec{algorithms:destination}.
As can be seen, Edge/Static shows non-negligible unfairness starting
at 4 clients, despite the overall load is rather limited.


\myssec{Network congestion scenario}{eval:tree-wsk}

\myfigeps[width=2.2in]{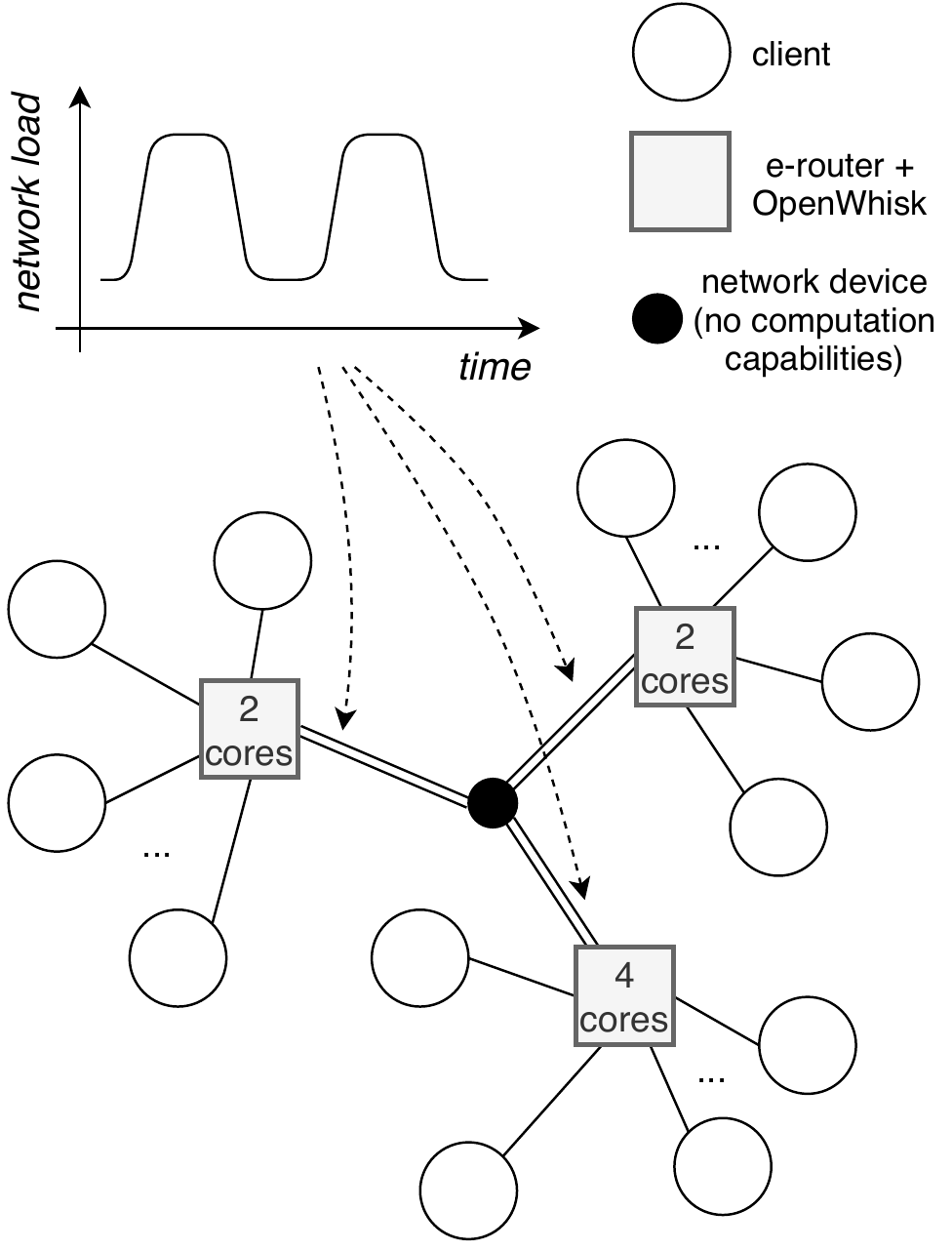}{%
Topology used in network congestion scenario (\rsec{eval:tree-wsk}).}

In this scenario we focus our attention on the mechanism of
congestion notification from the \ac{SDN} controller illustrated in
\rsec{algorithms:weight}.
To this purpose, we set up the network illustrated in \rfig{tree-topo},
with 10~Mb/s bandwidth / 2~ms latency links, and where we inject
periodically background TCP traffic in the links between the middle
node and the OpenWhisk executors, which are allocated 2-2-4 CPU
cores.
Only one link is interfered at a time.
We use CPU-intensive tasks only.
The clients, whose number increases from 5 to 30, are associated
randomly to one of the three e-routers, and they issue two requests
(10~kB each) per second, on average, uniformly distributed.

\myfigeps{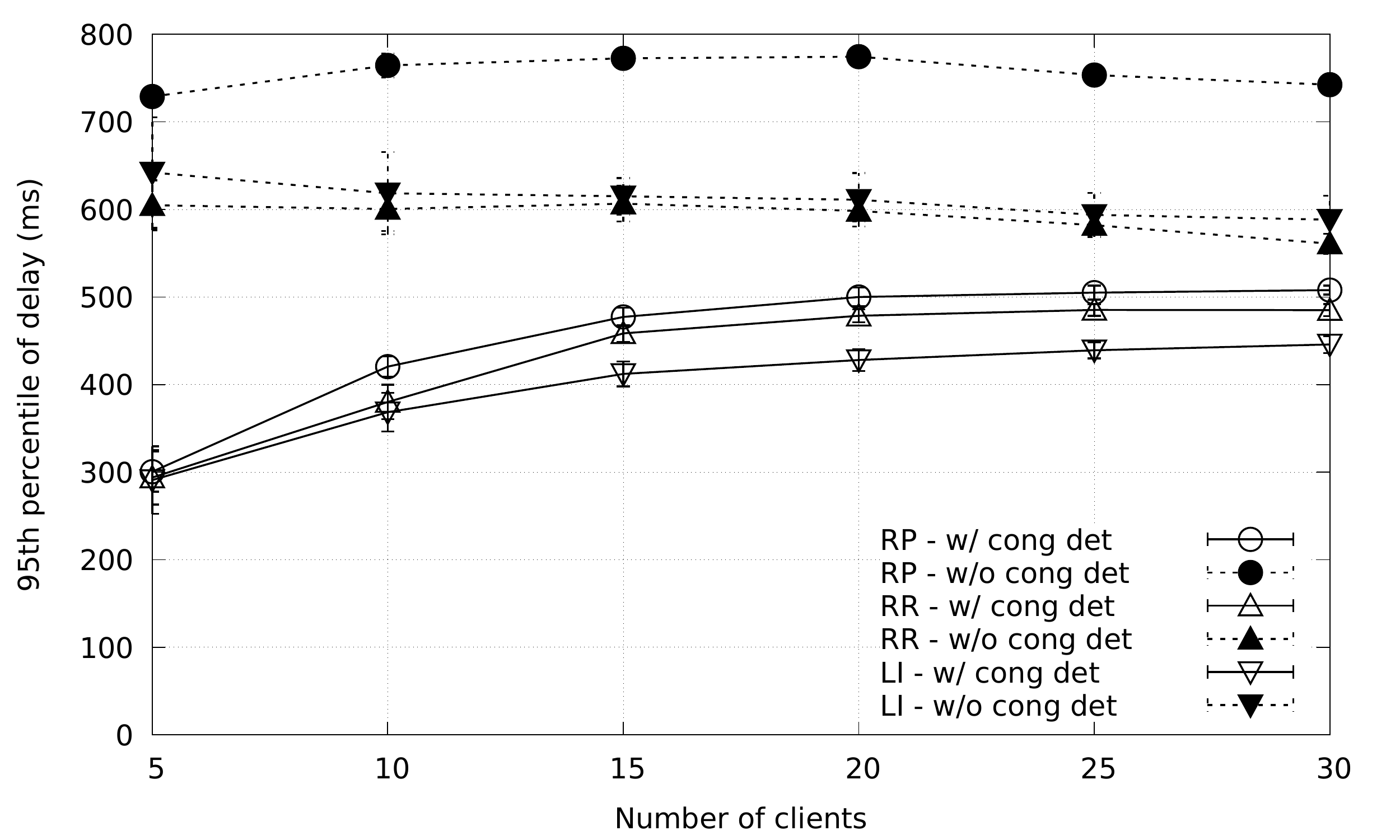}{%
Network congestion scenario:
95th percentile of delay vs.\ number of clients, with all destination
algorithms in \rsec{algorithms:destination}, and with/without
congestion detection in \req{weightupdate}.}

In \rfig{tree-wsk-out-095} we report the 95th percentile of delay.
As can be seen, without congestion notification, the 95th
percentile of delay does not depend on the load: whenever there is
congestion on a link, more than 5\% of lambda requests are affected
since the e-router on the congested link tries to execute them on 
executors after the path with background traffic.
This is avoided by enabling the congestion notification mechanism.
In this scenario, unlike that in \rsec{eval:grid}, LI exhibits
best performance, with RR being intermediate.
This is because LI suffers most from executors having an uneven
distribution of computational resources, which does not happen here.
Overall, RR is found to be more flexible in different conditions
than both RP and LI.



\myssec{Large scale scenario}{eval:large}

\myfigeps{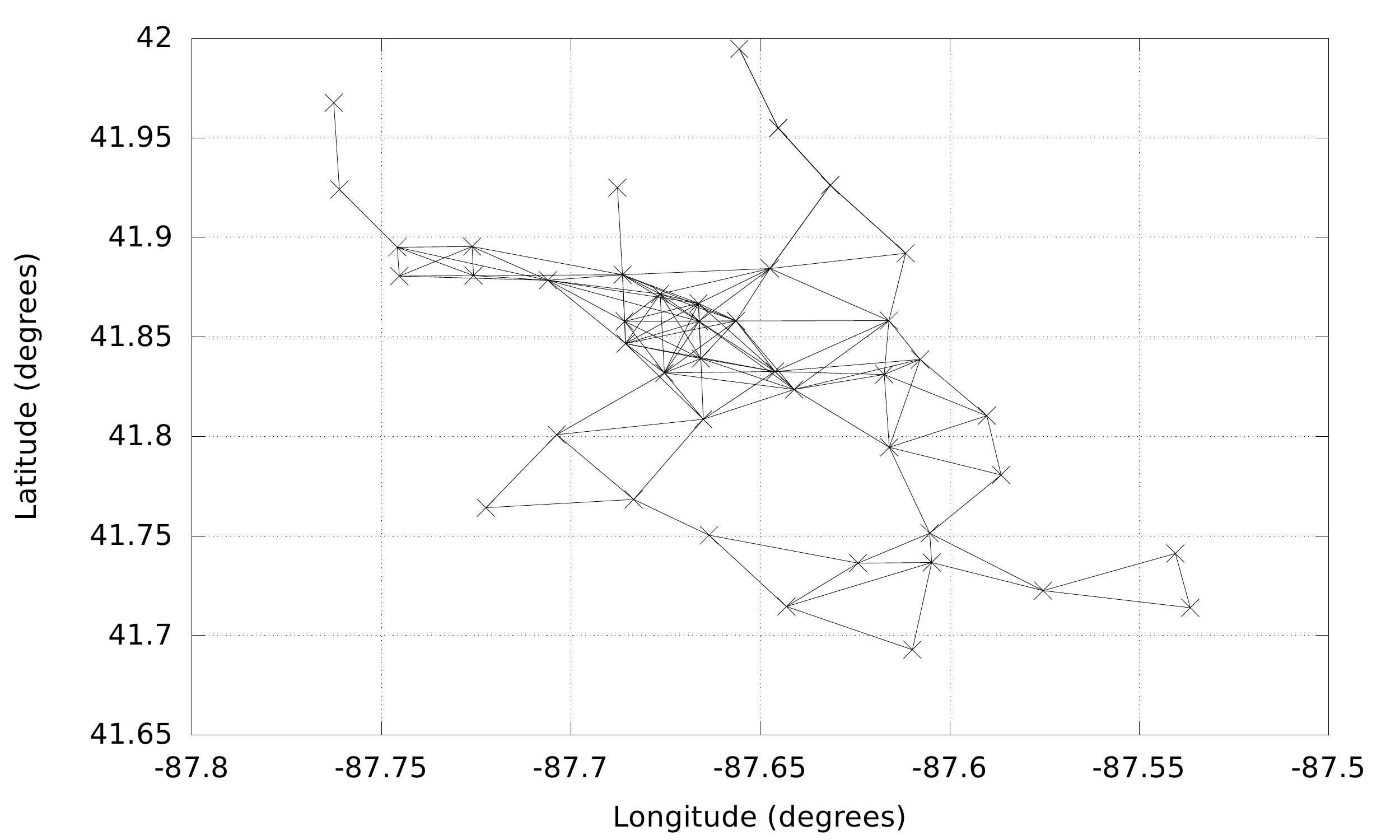}{Large scale scenario (\rsec{eval:large}): Network map.}

In this section we investigate the impact of using a two-tier
overlay, as described in~\rsec{scalability}, in a topology extracted
from a real \ac{IoT} network: Array of Things%
\footnote{\url{https://arrayofthings.github.io/}}, simplified by
collapsing nodes that are very close to one another.
The resulting network map is illustrated in~\rfig{chicago_topo}:
it consists of 45~nodes with a diameter of 11~hops.
All links have a 10~Mb/s capacity, with a 2~ms latency, emulating
an urban \ac{WMN}.
%
%
In this scenario we focus on the hierarchical forwarding scheme
in~\rsec{scalability}, hence we use emulated e-computers and clients
issue lambda requests with a fixed input size (equal to 5000~bytes,
corresponding to 8.3~ms processing time), so that the impact of
network transfer on the lambda response latency shows significantly.

We have assumed that all edge nodes host exactly
one e-computer offering the same lambda, whereas 10 e-routers are
dropped at the beginning of every run in random edge nodes.
We increased the number of clients from 30 to 120, and they also
are dropped at random locations at the beginning of every experiment.
During the experiment all clients continuously repeat the execution
of the same lambda request, thus the total number of lambda executions
is different for every run.
We compare a flat overlay to a two-tier overlay with the min-max
and min-avg policies defined in~\rsec{scalability:algo}.
Furthermore, only for evaluation purpose, we add a third policy
with a two-tier overlay, called \textit{random}: select a random
home e-router for every computer, thus emulating a topology-unaware
algorithm to realize the overlay, such as if using a \ac{DHT}, as
proposed in~\cite{8086146}.
The use of a two-tier overlay is found \textit{a posteriori} to
exhibit an average size of the e-tables equal to $8.4 \pm 3.5$,
while the size is always 45 with a flat overlay.
The destination selection algorithm is always RR, since it has been
shown above to provide the best performance compromise.

\myfigeps{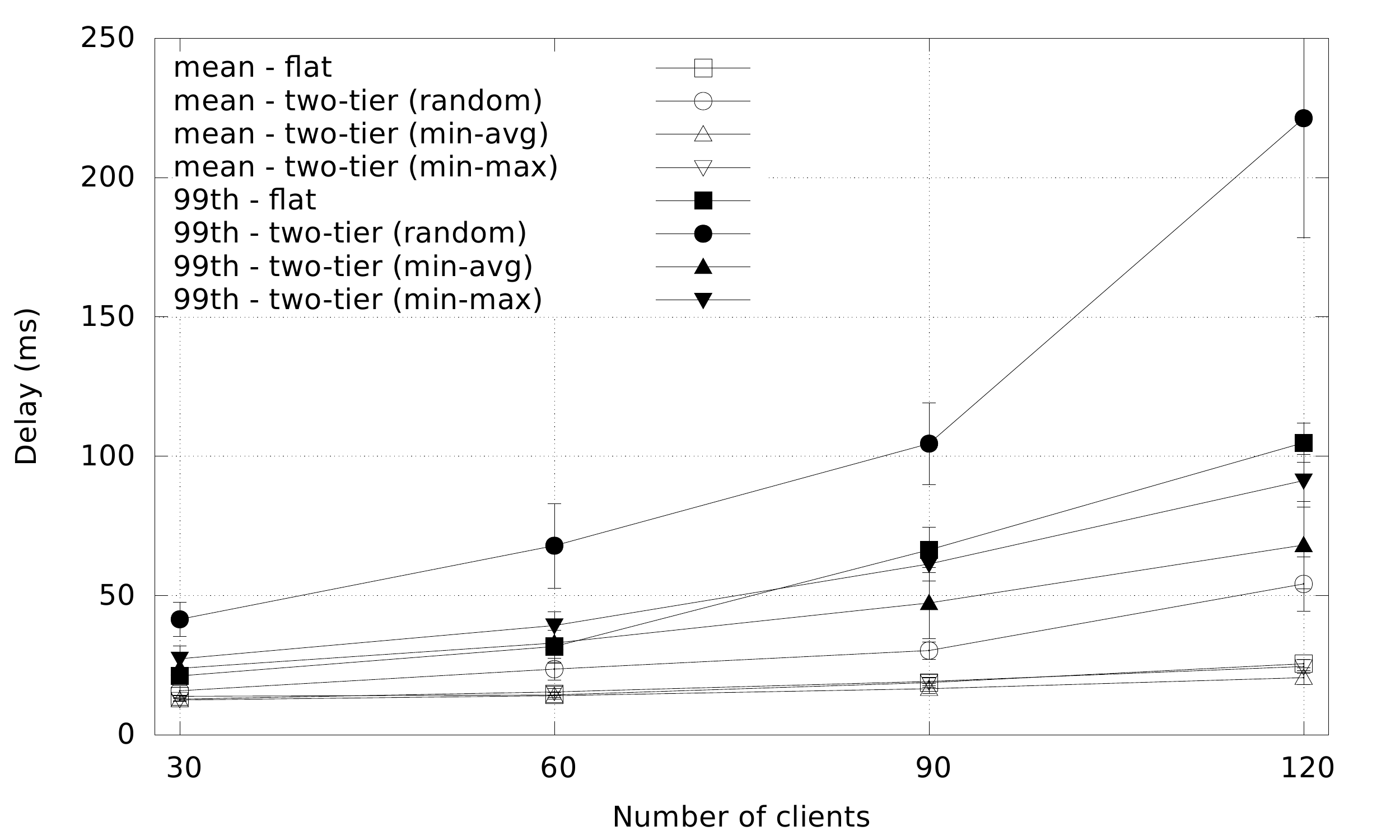}{%
Large scale scenario: Mean and 99th percentile of delay.}

In~\rfig{chicago-delay} we report the delay.
As can be seen, both the mean and the 99th percentile show the same
behavior, though more exacerbated in the latter, which is explained
in the following.
First, two-tier overlay (random) degrades the performance, even at
low loads, and more so as the network becomes overloaded.
This confirms our intuition in~\rsec{scalability} that the network
topology should be kept into account when determining the home
e-routers in a two-tier overlay, otherwise a performance penalty
must be paid because of the unnecessary long paths from the ingress
e-router to the home e-router and then to the e-computer for every
lambda request issued by a client.
Second, a flat overlay entails slightly smaller delays at low loads:
this is because it incurs less network transfer overhead, which is
the dominating factors.
However, as the number of clients increases, the e-computers become
gradually more loaded, hence the processing time becomes the primary
source of delay: in these conditions, a two-tier overlay, either
min-max or min-avg, is always beneficial in terms of delay, with
the latter (min-avg) yielding a 99th percentile of delay which is
about 40\% smaller than that with a flat overlay.
The reason is that with a two-tier overlay lambda execution tends
to reward proximity: remote e-computers are clustered and masked
as a single destination by their respective home e-routers, hence
probing is more lightweight and closer destinations are given a
little more than their ``fair'' share\footnote{This effect is
amplified if the topology offers a natural way to locate the home
e-routers, e.g., in a tree topology where e-routers are on any
intermediate level and the e-computers are on the leaves: such a
limit case is shown for completeness in the supplementary material.}

\myfigeps%
{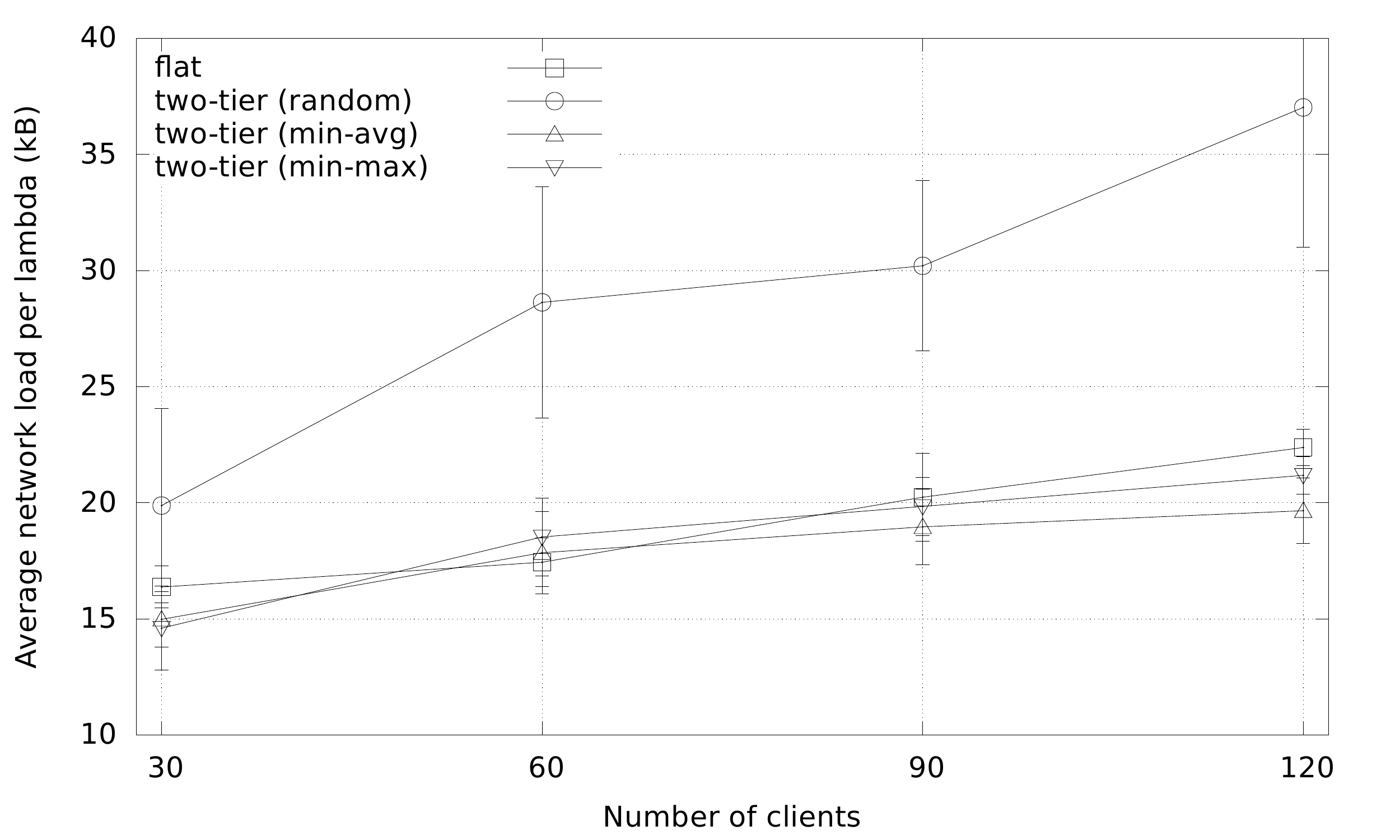}%
{Large scale scenario: Average network load per lambda request.}

Thus, a two-tier overlay is not only beneficial since it reduces
the size of the e-tables, but it also yields smaller delays.
This extremely positive result cannot be generalized to \textit{any}
possible environment: for instance, if the destinations closer to
an e-router are overloaded for any reasons, then a flat overlay may
utilize better resources that are far away but, in this case,
preferable.
However, we consider remarkable that a two-tier overlay, introduced
as a \textit{necessity} to reduce the size of e-tables for computational
reasons at the calculated risk of increasing network overhead, does
not in fact incur a penalty in terms of the latter in a realistic
\ac{IoT} network topology.
To stress this point, we report in~\rfig{chicago-tpt-lambda} a
direct measure of the average cost, in terms of network resources,
incurred by the execution of a single lambda.
This measure is defined as the ratio of the overall number of bytes
transmitted in the network by the total number of lambdas executed.
As can be seen, the relative performance of the different approaches,
in terms of this metric, is the same as in terms of the delay: the
two-tier with random policy incurs the highest overhead by far,
whereas the two-tier min-max and min-avg policies generally pay the
smallest network cost.
%

%% file: conclusions.tex
In this paper we have proposed an architecture for the realization
of serverless computing, which is of growing interest to \ac{IoT}
applications, in an edge network.
%
%
The clients send lambda function requests to the e-routers,
which forward them to the e-computers deemed to be most appropriate
at the moment.
%
%
The decision is taken based on weights local to every e-router,
which is however notified by the \ac{SDN} controller of congestion
events.
The overall solution is extremely lightweight and can be implemented
on devices with constrained resources, such as \ac{IoT} gateways,
especially when using a two-tier overlay option to reduce the
size of the local state.
We have designed three algorithms to select the destination of
lambda requests from clients, one of which, called RR, is proved
to guarantee both short- and long-term fairness.
%
%
%
%
We have validated extensively our contribution in different scenarios
via emulation experiments, also integrated with a widely-used open
source serverless framework (OpenWhisk).
Results have shown that our architecture makes it possible to handle
efficiently fast changing load and network conditions, in particular
the delay is comparable to that obtained in the optimistic case of
\added{a serverless platform with a distributed architecture}.
Furthermore, a two-tier overlay is effective in reducing the
computational needs while achieving the same or better performance
than a flat overlay.

In the future work we plan to study how the long-term allocation
of containers may benefit from real-time information provided by
the edge components.

%% file: biographies.tex
%

\begin{IEEEbiography}[{\includegraphics[width=1in,height=1.25in,clip,keepaspectratio]{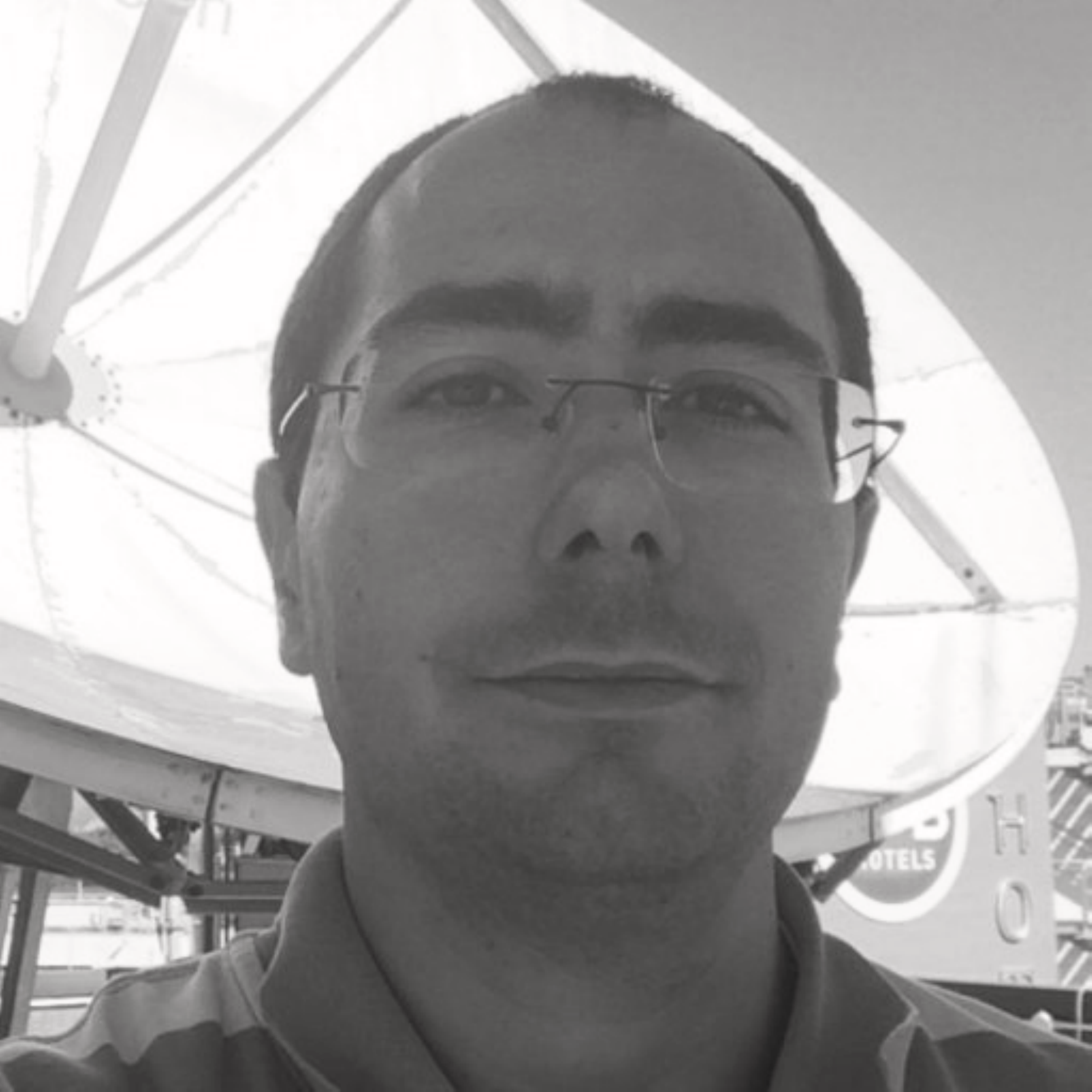}}]{Claudio Cicconetti}%
  \input{bios/claudio}
\end{IEEEbiography}

\begin{IEEEbiography}[{\includegraphics[width=1in,height=1.25in,clip,keepaspectratio]{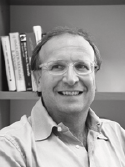}}]{Marco Conti}%
  \input{bios/marco}
\end{IEEEbiography}


\begin{IEEEbiography}[{\includegraphics[width=1in,height=1.25in,clip,keepaspectratio]{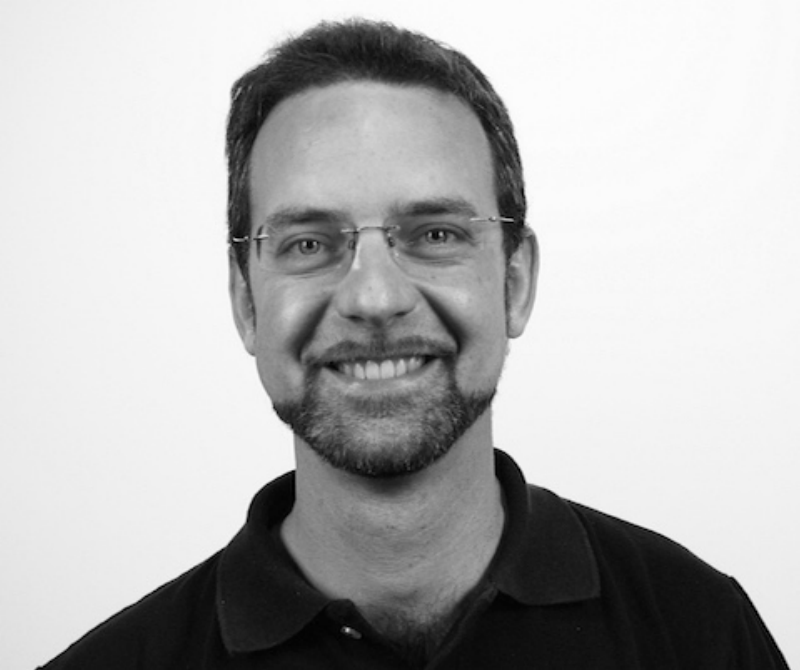}}]{Andrea Passarella}%
  \input{bios/andrea}
\end{IEEEbiography}



%% file: bios/claudio.tex
has a PhD in Information Engineering from the University of Pisa
(2007), where he also received his Laurea degree in Computer Science
Engineering.
He has been working in Intecs S.p.a. (Italy) from 2009 to 2013 as
an R\&D manager and in MBI S.r.l. (Italy) from 2014 to 2018 as a
principal software engineer.
He is now a researcher in the Ubiquitous Internet group of IIT-CNR
(Italy).
He has been involved in several international R\&D projects funded
by the European Commission and the European Space Agency.
%
%
%

%% file: bios/marco.tex
is the Director of IIT-CNR.
He was the coordinator of FET-open project ``Mobile Metropolitan
Ad hoc Network (MobileMAN)'' (2002-2005), and he has been the CNR
Principal Investigator (PI) in several projects funded by the
European Commission: FP6 FET HAGGLE (2006-2009), FP6 NEST MEMORY
(2007-2010), FP7 EU-MESH (2008-2010), FP7 FET SOCIALNETS (2008-2011),
FP7 FIRE project SCAMPI (2010-2013), FP7 FIRE EINS (2011-15) and
CNR Co-PI for the FP7 FET project RECOGNITION (2010-2013).
%
%
He has published in journals and conference proceedings more than
300 research papers
to design, modelling, and performance evaluation of computer and
communications systems.
%

%% file: bios/andrea.tex
is a Research Director at IIT-CNR and Head of the Ubiquitous Internet Group.
Before joining UI-IIT he was a Research Associate at the Computer
Laboratory of the University of Cambridge, UK.
%
%
He published 150+ papers in international journals and conferences,
receiving 4 best paper awards, including at IFIP Networking 2011
and IEEE WoWMoM 2013.
%
%
%
%
%
He was Chair/Co-Chair of several IEEE and ACM conferences/workshops
(including IFIP IoP 2016, ACM CHANTS 2014 and IEEE WoWMoM 2011 and 2019).
%
He is the founding Associate EiC of the Elsevier journal Online
Social Networks and Media (OSNEM), and Area Editor for the Elsevier
Pervasive and Mobile Computing Journal and Inderscience Intl.
Journal of Autonomous and Adaptive Communications Systems.
%
%
%

%% file: acronyms.tex
\begin{acronym}
  \acro{3GPP}{Third Generation Partnership Project}
  \acro{5G-PPP}{5G Public Private Partnership}
  \acro{AA}{Authentication and Authorization}
  \acro{API}{Application Programming Interface}
  \acro{AP}{Access Point}
  \acro{AR}{Augmented Reality}
  \acro{BGP}{Border Gateway Protocol}
  \acro{BSP}{Bulk Synchronous Parallel}
  \acro{BS}{Base Station}
  \acro{CDF}{Cumulative Distribution Function}
  \acro{CFS}{Customer Facing Service}
  \acro{CPU}{Central Processing Unit}
  \acro{DHT}{Distributed Hash Table}
  \acro{DNS}{Domain Name System}
  \acro{ETSI}{European Telecommunications Standards Institute}
  \acro{FCFS}{First Come First Serve}
  \acro{FSM}{Finite State Machine}
  \acro{FaaS}{Function as a Service}
  \acro{GPU}{Graphics Processing Unit}
  \acro{HTML}{HyperText Markup Language}
  \acro{HTTP}{Hyper-Text Transfer Protocol}
  \acro{ICN}{Information-Centric Networking}
  \acro{IETF}{Internet Engineering Task Force}
  \acro{IIoT}{Industrial Internet of Things}
  \acro{IPP}{Interrupted Poisson Process}
  \acro{IP}{Internet Protocol}
  \acro{ISG}{Industry Specification Group}
  \acro{ITS}{Intelligent Transportation System}
  \acro{ITU}{International Telecommunication Union}
  \acro{IT}{Information Technology}
  \acro{IaaS}{Infrastructure as a Service}
  \acro{IoT}{Internet of Things}
  \acro{JSON}{JavaScript Object Notation}
  \acro{LCM}{Life Cycle Management}
  \acro{LL}{Link Layer}
  \acro{LTE}{Long Term Evolution}
  \acro{MAC}{Medium Access Layer}
  \acro{MBWA}{Mobile Broadband Wireless Access}
  \acro{MCC}{Mobile Cloud Computing}
  \acro{MEC}{Multi-access Edge Computing}
  \acro{MEH}{Mobile Edge Host}
  \acro{MEPM}{Mobile Edge Platform Manager}
  \acro{MEP}{Mobile Edge Platform}
  \acro{ME}{Mobile Edge}
  \acro{ML}{Machine Learning}
  \acro{MNO}{Mobile Network Operator}
  \acro{NAT}{Network Address Translation}
  \acro{NFV}{Network Function Virtualization}
  \acro{NFaaS}{Named Function as a Service}
  \acro{OSPF}{Open Shortest Path First}
  \acro{OSS}{Operations Support System}
  \acro{OS}{Operating System}
  \acro{OWC}{OpenWhisk Controller}
  \acro{PMF}{Probability Mass Function}
  \acro{PU}{Processing Unit}
  \acro{PaaS}{Platform as a Service}
  \acro{PoA}{Point of Attachment}
  \acro{QoE}{Quality of Experience}
  \acro{QoS}{Quality of Service}
  \acro{RPC}{Remote Procedure Call}
  \acro{RR}{Round Robin}
  \acro{RSU}{Road Side Unit}
  \acro{SBC}{Single-Board Computer}
  \acro{SDN}{Software Defined Networking}
  \acro{SMP}{Symmetric Multiprocessing}
  \acro{SRPT}{Shortest Remaining Processing Time}
  \acro{STL}{Standard Template Library}
  \acro{SaaS}{Software as a Service}
  \acro{TCP}{Transmission Control Protocol}
  \acro{TSN}{Time-Sensitive Networking}
  \acro{UDP}{User Datagram Protocol}
  \acro{UE}{User Equipment}
  \acro{URI}{Uniform Resource Identifier}
  \acro{URL}{Uniform Resource Locator}
  \acro{UT}{User Terminal}
  \acro{VANET}{Vehicular Ad-hoc Network}
  \acro{VIM}{Virtual Infrastructure Manager}
  \acro{VM}{Virtual Machine}
  \acro{VNF}{Virtual Network Function}
  \acro{WLAN}{Wireless Local Area Network}
  \acro{WMN}{Wireless Mesh Network}
  \acro{WRR}{Weighted Round Robin}
  \acro{YAML}{YAML Ain't Markup Language}
\end{acronym}